\documentclass[preprint,12pt]{elsarticle}

\usepackage{amssymb}
\usepackage{amsmath}

\usepackage{graphicx} 
\usepackage{longtable}
\usepackage{makecell}
\usepackage{hyperref}
\usepackage{changepage} 

\date{April 2026}
\usepackage{enumitem}
\usepackage{soul}
\usepackage{xcolor}
\usepackage{color,colortbl}
\usepackage{adjustbox}
\usepackage{booktabs} 
\usepackage{multirow}
\usepackage{hyperref}
\usepackage{listings}
\usepackage{float}
\usepackage{url}
\usepackage{amssymb}

\usepackage{rotating}

\journal{Nuclear Physics B}

\begin{document}

\begin{frontmatter}



\title{
Protecting On-Device AI Inference: A Systematic Review of Attacks and Defence Mechanisms}


\author[Mitel]{Zisis Tsiatsikas\fnref{fn1}}
\author[aegean]{Alexandros Fakis}

\affiliation[Mitel]{
  organization={Mitel Networks},
  city={Athens},
  postcode={15124},
  country={Greece}
}

\fntext[fn1]{Disclaimer: The views and opinions expressed in this article are those of the author and do not necessarily reflect the official position of Mitel Networks. Mitel Networks cannot be held responsible for any use that may be made of the information contained in this article.}

\affiliation[aegean]{
  organization={Department of Information and Communication Systems Engineering, University of the Aegean},
  city={Samos},
  postcode={83200},
  country={Greece}
}

\author[jrc]{Georgios Karopoulos\fnref{fn2}}\ead{georgios.karopoulos@ec.europa.eu}

\affiliation[jrc]{
  organization={European Commission, Joint Research Centre (JRC)},
  city={Ispra},
  postcode={21027},
  country={Italy}
}

\fntext[fn2]{Corresponding author}

\author[efsa]{Vasileios Kouliaridis\fnref{fn3}}

\affiliation[efsa]{
  organization={European Food Safety Authority (EFSA)},
  city={Parma},
  postcode={43126},
  country={Italy}
}

\fntext[fn3]{Disclaimer: The views and opinions expressed in this article are those of the author and do not necessarily reflect the official position of the European Food Safety Authority (EFSA). EFSA cannot be held responsible for any use that may be made of the information contained in this article.}

\author[duth]{Marios Anagnostopoulos}

\affiliation[duth]{
  organization={Department of Electrical and Computer Engineering, Democritus University of Thrace},
  city={Xanthi},
  postcode={67100},
  country={Greece}
}
            
\begin{abstract}

The need for secure and private Artificial Intelligence (AI) and Machine Learning (ML) on edge and mobile devices has increased the necessity of protecting the architecture of these systems from threats to both security and privacy. With an ever-increasing number of pre-trained AI models being used on mobile platforms for client-side inference, there are rising concerns about the risks associated with the theft/extraction of AI models, adversarial attacks on AI models, and data breaches.
As a result of this trend, a variety of defence mechanisms have been proposed to protect against these threats. These include Trusted Execution Environments (TEEs), homomorphic encryption, obfuscation, and differential privacy, among others. 
However, current surveys largely focus on edge intelligence, which includes distributed training, and thus overlook security and privacy issues that are specific to on‑device AI inference. To the best of our knowledge, this paper presents the first comprehensive review of threats and corresponding defence mechanisms targeting on‑device inference.
Our results show that the attack and defence literature are unbalanced: approximately one quarter of the surveyed attack papers focus on Intellectual Property (IP) attacks, whereas half of the defence solutions tackle the same issue. More importantly, some attack categories have no defence paper associated to them, such as adversarial attacks that account for roughly one third of the attack literature. This asymmetry between known attacks and available mitigations highlights clear opportunities for future research on securing on-device AI inference. 

\end{abstract}


\end{frontmatter}



\section{Introduction} 

Over the last decade, we witness a rapid evolution in Artificial Intelligence (AI) technology, beginning with Machine Learning (ML) models, including k-Nearest Neighbors (k-NN), Random Forests, Support Vector Machines (SVM)~\cite{FARNAAZ2016213} to today’s Large Language Model (LLMs), which are equipped with billions of parameters and can be used to complete numerous complex tasks~\cite{zhang2024mmllmsrecentadvancesmultimodal}. Some examples of AI models currently leading the way, include LLaMA, Claude, and ChatGPT, each of which has been widely adopted by individuals and organisations, and has a major impact on how content is created~\cite{YAO2024100211}. Beyond text-based processing, modern models are also now capable of performing a wide variety of additional common tasks, such as image and object recognition, as well as transcription and summarisation of various forms of data; these are some of the cornerstones of the emerging Generative AI (GenAI) paradigm.
This rapid progress has created strong demand for AI‑enabled services across many industries~\cite{ESTEVEZ2025100796}.

Modern AI architectures depend heavily on cloud-based systems to provide the most efficient use of resources and reach maximum scalability~\cite{Adeboye2025GPU}. In addition to providing flexible usage to their clients (i.e., the ability to scale up or down), the pay-as-you-go subscription model also provides all of the ``under-the-hood'' details about how many requests will have access to a service and how AI models are deployed or upgraded. Additionally, the majority of consumers choose to go through cloud vendors because they provide a large variety of selection and provisioning options for each type of AI model and adhere to the most common security principles and frameworks, including ISO 27001~\cite{10.5555/1481582} and ISO 27017~\cite{ISO27017}. Although cloud-based AI services provide a high level of security to the user, they create significant barriers to entry for privacy-sensitive applications (e.g., healthcare and finance), where regulatory requirements typically limit the transfer of sensitive data from local devices to external servers~\cite{zhu2025fintechmeetsprivacysecuring}. These constraints prevent users from capitalising on cloud-based AI services and restrict the GenAI features available to users.

To satisfy security, privacy, latency, and bandwidth requirements, practitioners are increasingly deploying AI models that run entirely on the edge device.
This family of AI models covers a wide range of applications, from browser‑based LLM inference for privacy‑preserving text processing~\cite{WebLLM2024} to intelligent vehicle perception tasks, such as vehicle and pedestrian detection, passenger counting, and road‑anomaly sensing~\cite{s25103191}. 
The mobile-oriented AI ecosystem is expanding rapidly, supported by numerous public repositories that provide easily accessible open-source pre-trained models~\cite{PeaTMOSS}. 
These solutions are deliberately small, energy‑aware, and tuned to the computational limits of smartphones, wearables, and Internet of Things (IoT) nodes. 
Running inference locally gives users direct control over both the model and the data, reduces exposure to network‑based attacks, and narrows the attack surface to the device itself. 
Existing deployments are typically isolated either in software (e.g., WebAssembly modules~\cite{eynard2025wasmagents}) or in hardware enclaves, such as Trusted Execution Environments (TEEs)~\cite{7345265}. 
However, although isolation raises the bar for attackers, numerous studies have demonstrated techniques for bypassing these protections~\cite{9785622}.

\textbf{Our Contribution:} Several recent surveys have examined AI inference on edge and mobile platforms, yet only a few address security and privacy concerns explicitly~\cite{10601684, liu25, mukherjee2020}. Moreover, existing defence strategies for protecting AI models on edge devices remain fragmented, and no comprehensive framework exists for comparing them. To fill this gap we performed a systematic literature review of 71 peer‑reviewed works published between 2019 and 2025. To the best of our knowledge, this is the first study that focuses exclusively on security threats and corresponding defence mechanisms for on‑device edge AI.



Our contributions are threefold:
\begin{itemize}
\item \textbf{Survey assessment:} We analyse all existing review articles on edge and on-device AI inference dealing with security and privacy issues, identifying gaps in coverage. 

\item \textbf{Attack taxonomy:} We provide a taxonomy of the major attack families targeting on‑device AI, describe their inter‑relationships, and relate them to the current threat landscape. 

\item \textbf{Defence taxonomy and mapping:} Based on the attack taxonomy analysis, we offer a multidimensional mapping of state‑of‑the‑art defensive mechanisms, indicating which attack categories each method aims to mitigate, representative attack scenarios, and the impact of each defence on the confidentiality and integrity of on‑device AI models. 
\end{itemize}



The remainder of the paper is organised as follows. In the following section, we outline background concepts and the scope of this research. Section~\ref{sec:related} analyses relevant surveys from the literature related to AI inference on mobile and edge devices. In Section~\ref{sec:methodology}, we present the methodology applied for searching and selecting the literature analysed in this survey. Following that, Section~\ref{sec:attack} presents the attacks against on-device AI models, highlighting their dependencies and how they are interconnected. Next, Section~\ref{sec:defence} provides a thorough evaluation of defence solutions against edge AI deployment threats. Finally, in Section~\ref{sec:discussion}, we discuss our findings, whereas in Section~\ref{challenges} we analyse the challenges of edge AI, provide suggestions for addressing them, and suggest areas for future research. We conclude our paper in Section~\ref{sec:conclusions}.

\section{Background and scope}


AI deployments can be classified into two broad categories: cloud-based (or centralised) and edge intelligence~\cite{liu25}. 
In the more traditional centralised paradigm, all of the data that was either collected or processed by an edge device is sent to remote cloud servers for analysis. 
Edge-based AI, on the other hand, processes data on an edge device eliminating the need to transmit them to the cloud or a data centre. This approach reduces the attack surface, removes response delays, and conserves network resources. Edge intelligence training is typically performed through Federated Learning (FL), where each node updates a model locally; afterwards, the model parameters are exchanged with a coordinating server that aggregates them into a global model that is then transmitted back to all the nodes.

Edge devices refer to any number of devices with very limited capabilities in terms of computing capacity, memory, battery life, and storage ~\cite{dhar21,Wang_2025}. As a result of their limited capabilities, many AI applications cannot operate efficiently, especially when building and training AI models. Local processing and inference can take place on edge devices, so that the need for cloud-based communication is eliminated. This leads to several benefits, including greater security and confidentiality, and lower latency. There are drawbacks to deploying AI models on edge devices, including reduced scalability, greater maintenance requirements, and restrictions on the type and complexity of AI models that can be used. Several factors contribute to the performance of AI model training on edge devices, including hardware, software libraries, algorithms, and AI training theories; therefore, these must be given serious consideration in order to optimise the efficiency and effectiveness of on-device AI.

Edge-based AI can involve three primary activities: edge caching, edge training, and edge inference. Edge caching involves collecting and storing data at the edge. Data is generated or collected by edge devices in the form of environmental information; or the data may be provided over the Internet. Edge training is the distributed training of AI models at the edge. Training can occur at both edge devices and edge servers. Edge inference involves the operation of running an AI model on an edge device or server to perform inference and to provide results without communicating with the cloud. In this case, the AI model was previously trained elsewhere, potentially on a large server, and then compressed in order to increase its speed and operational efficiency. Because AI models run on the device, they are limited by the device's limited resources, such as computational capability, memory, and energy reserves ~\cite{Wang_2025}. On the other hand, operating AI models on-device can provide faster responses and greater user privacy since the data is processed locally. The focus of this survey will be on security and privacy issues associated with performing inference on edge devices and exclude work relating to edge caching and training.
 
\section{Related Work}
\label{sec:related}
Table~\ref{T:surveys} provides an overview of the existing surveys and state-of-the-art review papers dealing with the security and privacy of on-device and edge intelligence, covering the period from 2019 to 2025. The freeze date for our survey is December 2025. It should be noted that, although our paper focuses on on-device inference, we have intentionally adopted a more comprehensive approach to incorporate surveys addressing security and privacy in the domain of edge intelligence. This decision is motivated by the lack of more specialised surveys and the fact that on-device inference is a subset of edge intelligence. Consequently, solutions designed for edge intelligence could potentially be applicable to the on-device context as well.

\begin{table}

\scriptsize
\begin{center}

\begin{tabular}{|p{0.8cm}|p{0.7cm}|>{\centering\arraybackslash}p{1cm}|p{0.5cm}|>{\centering\arraybackslash}p{1.1cm}|p{3cm}|p{4cm}|} 
\hline
\textbf{Work} & \textbf{Year} &\textbf{Scope} &\textbf{AI} &\textbf{Security focus} & \textbf{Security Aspects}&\textbf{Prevention Mechanisms} \\ [0.5ex]
\hline
~\cite{10601684}&2025 & Edge AI&  ML, DL, FL, DRL  & Yes& Security issues (network attacks, confidentiality, integrity, intrusion) and privacy (data, location and model leakage)& Cryptography, lightweight aggregation, FL, blockchain, TEE, clustering-based methods, differential privacy, lightweight cryptography, secure multi-party
computation, ML-based methods\\ 
\hline
~\cite{Wang_2025}&2025 & On-device AI&  ML, DL, RL, TL  &No&Data protection, compliance and security attacks& Anonymisation, homomorphic encryption, TEE, multi-party computation, data governance, FL \\ 
\hline
~\cite{liu25}&2025 & Edge AI & Gen &Yes  & Data and model privacy & Encryption (homomorphic, spatial transformation, secure multi-party computation), perturbation (differential privacy, additive perturbation, multiplicative perturbation), anonymisation (k-anonymity, l-diversity, t-closeness), anomaly detection, model tolerance \\ 
\hline
~\cite{huckelberry2024}&2024 & TinyML&  ML, DL & Yes& Hardware (side-channel, memory extraction/flashing, fault injection), Software (communication, model updates), ML Model (adversarial examples, model extraction, backdoor, model inversion) & Hardware: obfuscating, masking, voltage regulators, secure boot, RDP/RCROP/WRP, ECC, PVD\newline
Software: TLS, AES, ECDH, RIOT-ML
\\
\hline
~\cite{9596610}& 2021 & Edge AI &   ML, DL, FL&No & Data leakage, data and model poisoning & FL, credit system, secure aggregation, averaging aggregation, homomorphic encryption, differential privacy\\ 
\hline
~\cite{Murshed_2021}&2021 & Edge AI & ML, DL & No  & Training data privacy, user privacy, fraud detection & Cryptography, differential privacy, blockchain, sensitive data removal\\ 
\hline
~\cite{Deng_2020}&2020 & Edge AI & DNN & No &Privacy (data leak, model updates) & Blockchain-based FL, game theoretical approaches\\ 
\hline
~\cite{shi2020communicationefficientedgeaialgorithms}&2020& Edge AI & Gen &No & Data and model privacy & Data and model partitioning \\ 
\hline
~\cite{mukherjee2020}& 2020 & Edge AI &  Gen &  Yes  & Dataset poisoning, privacy attacks, attacks on learning agents & Secure data aggregation, anonymity, access control  \\ 
\hline
~\cite{8763885}&2019& Edge AI & DL & No  & Privacy & Perturbation, secure computation (homomorphic encryption, secure multiparty computation)\\ 
\hline
\end{tabular}

\end{center}
\caption{Related survey papers (ML: machine learning, DL: deep learning, FL: federated learning, DRL: deep reinforcement learning, RL: reinforcement learning, TL: transfer learning, Gen: artificial intelligence in general).}
\label{T:surveys}
\end{table}

Aouedi et al.~\cite{10601684}, conduct a survey on Intelligent IoT (IIoT) and focused on adding intelligence to edge devices with integrated local AI models. This work considers different security aspects, such as network attacks (including DoS, poisoning, and adversarial attacks), confidentiality (including confidentiality and access control), integrity, intrusion detection and prevention. It also reviews privacy issues, such as data, location, and model leakage. Different IoT applications are considered, including smart healthcare, smart cities, smart transportation, and smart industry. The survey also analyses challenges of IIoT related to security and privacy, including resource management, learning model design, fairness, and economic issues. Finally, the survey covers different aspects of edge intelligence, including interactions with the cloud. Our study, on the other hand, is focused on analysing the security and privacy of on-device inference.

Wang et al.~\cite{Wang_2025} survey on-device AI models, and their analysis focuses on real-world applications, technical challenges, performance enhancements, and future trends. The examined use-cases of on-device AI models are smartphones and similar mobile devices, IoT, edge computing, intelligent transportation systems, and medical devices. In the domain of security and privacy, this research only briefly examines data protection, compliance, and security attacks, concluding that more robust data protection mechanisms are required.

Liu et al.~\cite{liu25} analyse the privacy issues in edge intelligence and review the protection mechanisms, challenges, and future research directions. The work focuses on two main attack categories: (i) data-privacy attacks, such as inference attacks where attackers try to acquire private information through parameter updates, and (ii) model-privacy attacks, such as backdoor attacks where malicious content is inserted into the model training process. 
For data-privacy attacks, the existing privacy protection methods are categorised into: (i) encryption, comprising homomorphic encryption, spatial transformation, and secure multi-party computation, (ii) perturbation, comprising differential privacy, additive perturbation, and multiplicative perturbation, and (iii) anonymisation, comprising k-anonymity, l-diversity, and t-closeness. The authors state that all methods have their advantages and disadvantages, such as high computational overhead and high security for encryption or low overhead and low security for perturbation, proposing a hybrid approach as potential solution.
For model-privacy attacks, the examined defences are classified into: (i) anomaly detection, where abnormal behaviour is identified when providing data input or performing model training, and (ii) model tolerance, where the model maintains its operation even when an attacker successfully tampers with the model. Bear in mind that each defence category has its own merits with anomaly detection being more resource intensive. A hybrid approach is also feasible by combining both strategies. Lastly, the authors discuss challenges and future research directions for both categories of attacks, including balancing protection and computational overhead, emerging technologies, like machine unlearning, and cross-layer defence systems.
Overall, ~\cite{liu25} provides a broad view of security and privacy issues across edge devices and cloud‑based services. However, only the first part of that work, which examines inference attacks, directly overlaps with the scope of our research; in contrast, the present survey is dedicated to on‑device inference. To this end, it not only surveys a more extensive set of attack vectors but also systematically reviews the corresponding defence techniques.

Huckelberry et al.~\cite{huckelberry2024} provide the first comprehensive security survey for TinyML systems, which deploy ML on resource-constrained microcontrollers with limited RAM and low-speed processors. The survey examines hardware attacks (side-channel, memory extraction, fault injection), software vulnerabilities (communication, model updates), and ML-specific attacks (adversarial examples, model extraction, backdoor, model inversion). A key finding is that robust countermeasures, such as masking, obfuscation, and TLS, are computationally infeasible for TinyML, whereas lightweight solutions, like adversarial training and hardware-level protection (secure boot, RDP/RCROP/WRP), remain viable. 
This survey focuses exclusively on TinyML microcontrollers and omits to cover the broader on-device AI landscape, including smartphones and IoT devices. On the contrary, our survey focuses on the broader on-device AI ecosystem, where the available computational resources substantially enable thorough security analysis and the deployment of robust defence mechanisms.

Xu et al.~\cite{9596610} consider the security and privacy of edge intelligence in two main directions: (i) reviewing security and privacy issues of edge training, and (ii) analysing future directions and challenges related to security and privacy. Regarding edge training, they focus on FL aggregation and examine leakage of sensitive data from updates, as well as data and model poisoning; they also review proposed solutions for these issues. The main open challenges identified include data leakage in edge intelligence in general and encryption in edge devices, given its intensive computation requirements.
Compared to our survey, this work briefly investigates security and privacy issues of edge intelligence and considers only the edge training phase.

Murshed et al.~\cite{Murshed_2021} study ML on edge devices, including ML-based applications and frameworks on the edge. Concerning security and privacy, they consider privacy protection of sensitive training data both from local devices and central aggregation servers, using cryptography, differential privacy, and blockchain as countermeasures. They also examine user data privacy and fraud detection where ML tools are used to authenticate data validity and detect false data.
This survey focuses on edge intelligence and only briefly examines security and privacy issues.

Deng et al.~\cite{Deng_2020} classify the edge intelligence concept into two categories: (i) intelligence-enabled edge computing, that is, solving edge computing issues with AI, and (ii) AI on the edge, namely, building AI models on the edge. In this context, privacy has a universal role covering both dimensions. More specifically, this work considers data privacy, and the prevention mechanisms surveyed include blockchain-based FL and game theoretical approaches.
On the other hand, this paper does not extensively cover  security and is rather focused on Quality of Experience (QoE) of edge intelligence, which treats security and privacy as one of the QoE aspects.

Shi et al.~\cite{shi2020communicationefficientedgeaialgorithms} investigate the advances in various techniques that assist in overcoming the heavy communication overheads incurred by edge AI. One of the general challenges identified for edge AI is the security and privacy of AI services. Subsequently, they deal with edge training systems and review the literature on data and model partitioning for privacy protection.
Regarding security and privacy, this survey provides limited analysis, focusing on data and model privacy.

Mukherjee et al.~\cite{mukherjee2020} briefly survey security and privacy issues in intelligent edge computing where AI is used to improve network management. The authors focus on the various attacks targeting edge intelligence and review open challenges and future directions. The main attacks examined are: (i) data poisoning to influence training and make the model take incorrect decisions, (ii) privacy compromise by intercepting or inferring private information that is exchanged among communication nodes, and (iii) attacks on the edge learning process by inserting a hidden attack vector into one learning agent who then infects all other agents by interacting with each other. Among the solutions discussed are secure data aggregation, anonymity, and access control. Regarding open issues that need further consideration, the authors mention security in FL, ubiquitous authentication of mobile devices, anonymity, and privacy.
This work considers AI on the edge from a different perspective than the rest of related papers by focusing on improving the utilisation of the edge computing system. Nevertheless, it is included in the discussion since the security and privacy issues are similar regardless of the AI application.

Chen and Ran~\cite{8763885} discuss key applications in which deep learning is used at the network edge. They analyse the benefits of deep learning on edge devices in terms of privacy, as well as the privacy challenges of such an approach. On the one hand, by moving the data analysis closer to the data source on edge devices, the exposure of potentially sensitive data on the public Internet is avoided. However, even in this case privacy is also seen as a challenge as some data still need to be shared among edge devices and/or the cloud. The solutions examined in this review are two privacy-preserving inference methods: adding noise to data and secure computation using cryptographic techniques, such as homomorphic encryption and secure multiparty computation.
This paper addresses only privacy concerns in the edge intelligence domain; however, it examines these issues in limited depth, as privacy is considered one of the challenges.

Upon reviewing existing research, we identified a notable gap in the study of security and privacy issues specifically related to on-device AI inference. 
Most recent work concentrates on edge intelligence, which, although related, covers a wider range of applications than on‑device inference. Making this distinction is important because some security threats that are relevant for edge intelligence do not apply to on‑device settings. For example, an attacker might attempt to eavesdrop on data exchanged among edge systems during collaborative training, for example, when using FL; such communication is absent when a model runs only on a single device.
The opposite is also true: certain threats are easier to realise in on‑device scenarios because of their different characteristics. A clear case is the theft of the entire AI model. An attacker who obtains physical access or privileged software access to a device can copy the complete model from its storage. In FL, the model is split among many participants and protected by aggregation protocols, making full‑model extraction much harder.
Additionally, many existing surveys address security and privacy only as part of a broader overview of edge AI, lacking detailed analysis and specialised focus. To date, there is only one survey on on-device AI that, however, does not concentrate on security, and three that focus on security but in the context of edge intelligence. Therefore, there is a lack of comprehensive surveys dedicated to the security of on-device inference. Our work addresses this gap by presenting the first, as per the authors' knowledge, detailed survey of security attacks and defences, specifically for on-device AI inference.


\section{Methodology} 
\label{sec:methodology}

This section describes the process used to identify, select, and classify the literature included in this survey. We follow a structured search and selection methodology adapted from established guidelines for systematic reviews of the literature in software engineering~\cite{kitchenham2007guidelines}.

\subsection{Search Approach}

A systematic search is conducted using the Scopus database to identify peer-reviewed journal articles and conference proceedings from major publishers, including IEEE, ACM, Elsevier, and Springer. This review encompasses research works published from January 2019 to December 2025, with the later serving as the cut-off date. The search is performed utilising two complementary queries that target attack-focused and defence-focused works, respectively. These two queries provide an overall balance of both sides of the security spectrum.

For attack-related literature, the following query string is applied:

\begin{lstlisting}[caption={Attack-focused search query.}]
TITLE-ABS-KEY(
  ("on-device inference" OR "edge inference"
   OR "on-device model" OR "mobile AI" OR "TinyML")
  AND ("machine learning" OR "deep learning"
       OR "neural network")
  AND ("attack" OR "adversarial" OR "model extraction"
       OR "model stealing" OR "side channel"
       OR "membership inference" OR "model inversion"
       OR "backdoor" OR "model theft")
)
AND PUBYEAR > 2018
AND (LIMIT-TO(DOCTYPE,"ar") OR LIMIT-TO(DOCTYPE,"cp"))
\end{lstlisting}

For defence-related literature, the query string was:

\begin{lstlisting}[caption={Defence-focused search query.}]
TITLE-ABS-KEY(
  ("on-device inference" OR "edge inference"
   OR "on-device model" OR "mobile AI" OR "TinyML")
  AND ("machine learning" OR "deep learning"
       OR "neural network")
  AND ("defense" OR "defence" OR "TEE"
       OR "trusted execution" OR "obfuscation"
       OR "model protection" OR "secure inference"
       OR "privacy-preserving" OR "differential privacy")
)
AND PUBYEAR > 2018
AND (LIMIT-TO(DOCTYPE,"ar") OR LIMIT-TO(DOCTYPE,"cp"))
\end{lstlisting}

The attack query returned 192 results, while the defence query returned 80 results. After removing duplicated results, the combined pool comprised 260 unique publications.

\subsection{Selection Criteria}

Following, each retrieved record is screened against four sets of criteria, applied sequentially. The general criteria are listed in Table~\ref{tab:general_criteria}, the inclusion criteria in Table~\ref{tab:inclusion_criteria}, the exclusion criteria in Table~\ref{tab:exclusion_criteria}, and the quality assessment criteria in Table~\ref{tab:quality_criteria}.

\begin{table}[b]
\centering
\small
\renewcommand{\arraystretch}{1.3}
\caption{General criteria.}
\label{tab:general_criteria}
\begin{tabular}{|c|p{0.78\columnwidth}|}
\hline
\textbf{No.} & \textbf{Criterion} \\
\hline
G1 & The paper is published between January 2019 and December 2025. \\
\hline
G2 & The paper is published in a peer-reviewed journal or peer-reviewed conference proceedings. \\
\hline
G3 & The paper addresses security or privacy of AI/ML models deployed on mobile, edge, or IoT devices. \\
\hline
\end{tabular}
\end{table}

\begin{table}
\centering
\small
\renewcommand{\arraystretch}{1.3}
\caption{Inclusion criteria.}
\label{tab:inclusion_criteria}
\begin{tabular}{|c|p{0.78\columnwidth}|}
\hline
\textbf{No.} & \textbf{Criterion} \\
\hline
I1 & The paper is written in English. \\
\hline
I2 & The paper targets the inference phase of ML model execution on a resource-constrained end-user device (mobile, edge, or IoT). \\
\hline
I3 & The paper presents a concrete technical contribution (attack method, defence mechanism, or empirical analysis) with experimental evaluation. \\
\hline
\end{tabular}
\end{table}

\begin{table}
\centering
\small
\renewcommand{\arraystretch}{1.3}
\caption{Exclusion criteria.}
\label{tab:exclusion_criteria}
\begin{tabular}{|c|p{0.78\columnwidth}|}
\hline
\textbf{No.} & \textbf{Criterion} \\
\hline
E1 & The paper focuses exclusively on cloud-based AI infrastructure or server-side processing. \\
\hline
E2 & The paper addresses federated learning training scenarios or distributed model training. \\
\hline
E3 & The paper does not present a primary technical contribution, but rather surveys or synthesises existing work. \\
\hline
E4 & The paper presents no empirical evaluation or prototype implementation. \\
\hline
E5 & The paper is a duplicate or an earlier version of an already-included work. \\
\hline
\end{tabular}
\end{table}

\begin{table}
\centering
\small
\renewcommand{\arraystretch}{1.3}
\caption{Quality assessment criteria.}
\label{tab:quality_criteria}
\begin{tabular}{|c|p{0.78\columnwidth}|}
\hline
\textbf{No.} & \textbf{Criterion} \\
\hline
Q1 & The paper clearly defines the threat model or defence objective. \\
\hline
Q2 & The proposed method or attack is described with sufficient technical details for replication. \\
\hline
Q3 & The evaluation is conducted on a real or widely accepted benchmark system, dataset, or device platform. \\
\hline
Q4 & The results are quantitatively reported and support the claims made by the authors. \\
\hline
Q5 & The attack or defence presented in the paper is sufficiently specific to be classifiable within a defined threat or defence category, rather than addressing on-device AI security only tangentially. \\
\hline
\end{tabular}
\end{table}

Overall, an article is included if it satisfied all general and inclusion criteria, met all quality assessment standards, and did not trigger any exclusion criteria. Conversely, a paper is excluded if it failed at least one inclusion criterion or satisfied at least one exclusion criterion. Furthermore, any publication identified as survey or review paper during the screening is excluded from the primary corpus but included in Section~\ref{sec:related}.

\subsection{Selection Process and Results}

The process for the literature review is carried out in three stages. In the first stage, a title and abstract screening is performed on the total of the 260 studies identified in the database search, reducing the number of potential studies by nearly 65\%. This results in a total of 90 studies that are then selected as candidates for inclusion. The second stage includes a full-text review against exclusion and quality criteria, which narrows the corpus even further. In the third stage, the remaining papers are organised by their respective focus areas, namely, attack-focused or defence-focused, which results in a final corpus of 61 studies. Of this pool, 17 of the papers deal with attack techniques, while the rest 44 address defensive countermeasures. In detail, the findings of each stage are summarised in Table \ref{tab:methodology}.

\begin{table}
\centering
\small
\renewcommand{\arraystretch}{1.3}
\caption{Literature selection summary.}
\label{tab:methodology}
\begin{tabular}{|l|r|}
\hline
\textbf{Stage} & \textbf{Records} \\
\hline
Attack search results (Scopus) & 192 \\
Defence search results (Scopus) & 80 \\
Combined pool (after deduplication) & 260 \\
\hline
Excluded after title/abstract screening & 170 \\
Excluded after full-text review & 29 \\
\hline
\textbf{Attack papers included} & \textbf{17} \\
\textbf{Defence papers included} & \textbf{44} \\
\textbf{Total included} & \textbf{61} \\
\hline
\end{tabular}
\end{table}

\subsection{Data Extraction and Classification}

For the analysis and evaluation of the surveyed publications, the following information is extracted from each of the included studies: primary attack/defence method (i.e., threat vector or countermeasure), target device, type of the ML model utilised, study design, and performance metrics. The attack studies are then categorised using the taxonomy defined in Section \ref{sec:attack}, whereas the defence studies are classified based on the respective taxonomy of Section \ref{sec:defence}.

\section{Attack Techniques and Exploitation Methods}
\label{sec:attack}

This section systematically reviews the attack landscape that targets machine learning models deployed on mobile and edge devices. We analyse the evolution of attack types and their impact on real-world mobile applications. Recently, a growing body of research has demonstrated a variety of potential vulnerability vectors and exploitation methods. In order to comprehensively analyse and compare the threat vectors discussed in this paper, we apply a multi-dimensional attack taxonomy which categorises attack types based on their objectives and exploitation techniques. The comprehensive taxonomy is shown in Table \ref{Tab:attacks}, including the specific attack vector and its resultant impact on mobile and edge AI systems. 

It should be noted that, for the purposes of this taxonomy, papers are characterised by their operative mechanism, that is, the primary technical method, rather than their downstream objective. Several works in the current survey combine techniques from multiple attack categories. In such cases, the categorisation is performed by the novel contribution of each work. For instance, a paper that reverse engineers a compiled model and uses the resulting model to launch adversarial perturbations, is classified under Intellectual Property, as opposed to Adversarial Attacks, since the extraction mechanism is the paper's core contribution. This principle results in a consistent categorisation, which avoids inconsistencies due to overlap in attack surfaces.

\begin{table}

\scriptsize  
\renewcommand{\arraystretch}{0.5}  
\setlength{\extrarowheight}{2pt}
\begin{tabular}{|p{2cm}|p{5.8cm}|p{5cm}|}
\hline
\textbf{Category} & \textbf{Examples} & \textbf{Impact} \\ \hline

\textbf{Physical} 
& \begin{itemize}[leftmargin=*, noitemsep, topsep=2pt, parsep=1pt, partopsep=0pt]
    \item Side Channel Attacks (Cache Timing, Power Analysis, GPU/TEE Profiling, Branch Prediction/Speculative Execution)
    \item Runtime Profiling Attacks
    \item Physical Memory Attacks (Memory Leakage, TEE Overlap, Physical Access)
    \item Malicious Interrupts
\end{itemize}
& Hardware-level compromise enabling extraction of model weights or sensitive data; bypasses system protections; threatens TEEs and resource-constrained devices \\ \hline

\textbf{Intellectual property} 
& \begin{itemize}[leftmargin=*, noitemsep, topsep=2pt, parsep=1pt, partopsep=0pt]
    \item Model Stealing/Extraction:
    \begin{itemize}[leftmargin=1em, noitemsep, topsep=1pt]
        \item Weight Theft
        \item Architecture Extraction/Reverse Engineering
        \item Black-Box Model Stealing
        \item Intermediate Feature Leakage
        \item Retraining Attacks (Freeze-and-Retrain)
        \item Pretrained-Model Side Information
    \end{itemize}
    \item Illegal Copy/Piracy
    \item Integer Overflow Exploits
    \item IP Violations
\end{itemize}
& Loss of proprietary assets; enables cloning of deployed models; reduces competitiveness and security of on-device ML systems \\ \hline

\textbf{Adversarial} 
& \begin{itemize}[leftmargin=*, noitemsep, topsep=2pt, parsep=1pt, partopsep=0pt]
    \item White-Box Attacks (FGSM, PGD, C\&W)
    \item Black-Box Attacks (Transfer-Based, Query-Based)
    \item Evasion Attacks
    \item Targeted/Untargeted Adversarial Examples
    \item Poisoning (Backdoor)
\end{itemize}
& Model misclassification causing failure of safety-critical applications; attackers can bypass authentication, detection, or decision systems \\ \hline

\textbf{Data privacy} 
&
    \begin{itemize}[leftmargin=*, noitemsep, topsep=2pt, parsep=1pt, partopsep=0pt]
    \item Membership Inference Attacks
    \item Model Inversion Attacks
    \item Training Data Exposure (Gradient Leakage, Graph Link-Stealing)
    \item Intermediate Feature Leakage
    \item Input Data Leakage
\end{itemize}
& Leakage of personal or sensitive information (images, speech, medical records); privacy compromise of users and device owners \\ \hline

\textbf{Integrity} 
& \begin{itemize}[leftmargin=*, noitemsep, topsep=2pt, parsep=1pt, partopsep=0pt]
    \item Model Tampering
    \item Backdoor Attacks
    \item Payload Injection
    \item Unauthorised Access (Forged Licences)
    \item Inference Integrity Attacks
    \item Preprocessing Pipeline Tampering
\end{itemize}
& Corruption of on-device model behaviour; insertion of malicious functionality; unauthorised model use; reliability degradation; compromised inference results \\ \hline

\textbf{Availability} 
& \begin{itemize}[leftmargin=*, noitemsep, topsep=2pt, parsep=1pt, partopsep=0pt]
    \item Resource Exhaustion (DoS) (Energy-Latency Attack)

\end{itemize}
&  Disruption or degradation of on-device AI services through resource exhaustion or performance manipulation, preventing timely or reliable inference without necessarily altering model outputs.\\ \hline

\end{tabular}
\caption{Attack taxonomy for on-device AI systems, showing example attacks in each category and their impacts on mobile and edge devices.}
\label{Tab:attacks}
\end{table}

\subsection{Taxonomy}
\label{Threat}

This section presents our attack taxonomy for machine learning systems on devices, which provides a basis for understanding the defensive mechanisms discussed in Section~\ref{sec:defence}. We organise attacks into six categories according to their targets and objectives: physical, intellectual property, adversarial, data privacy,integrity, and availability attacks.

\textit{Physical attacks} focus on vulnerabilities in the hardware and/or micro-architectural implementations of computer systems and circumvent all other software-based protection mechanisms. In most cases, the attack involves side-channel exploitation, where attackers can obtain sensitive information from a system by monitoring its physical execution characteristics. These characteristics include timing variations, electromagnetic emissions, and runtime profiling data. Unlike traditional software-based attacks, physical attacks utilise side-channel leakages that are specific to the hardware layer. Successful physical attacks typically require physical access, either co-location on the same device or physical proximity, to achieve high-impact objectives, such as extracting machine learning models from on-device AI deployments.

\textit{Intellectual property attacks} represent one of the most prevalent threats to on-device ML models. In this case, an attacker attempts to steal proprietary models using extraction and reverse engineering techniques. This type of attacks can potentially extract model architectures, parameters, weights, and computational structures, resulting in the theft of intellectual property, unauthorised model cloning, while also allowing subsequent attacks. Direct extraction attacks target accessible model files stored on devices, whereas more sophisticated attacks focus on reverse engineering compiled model artifacts. One of the main reasons that reverse engineering compiled model artifacts is so challenging is due to the fact that many mobile ML frameworks, such as TensorFlow Lite, are able to compile models into inference-only versions of the model, lacking back-propagation capability. This compilation removes readability, thus forcing the attacker to recreate the original model from optimised binary code.

\textit{Adversarial attacks} represent a type of manipulation that may result in incorrect predictions or misclassification of the model's output. There is considerable variability with respect to the threat models, the access levels required for an adversary, and how the model may be impacted by these attacks. White box attacks provide an adversary with full knowledge of the model's architecture, parameters, and training methodology. Black box attacks do not provide the adversary with any information about the target system, however, they may employ techniques to identify potential vulnerabilities based on transfer attacks or query-based methods. These types of attacks present significant issues with the deployment of on-device AI due to the limited resources available to perform robust input validation and detect the presence of an attack.

\textit{Data privacy attacks} extract sensitive information from training data or identify dataset members. These attacks compromise training data confidentiality, whereas adversarial attacks target model accuracy. Model Inversion reconstructs training samples from model outputs. Membership Inference determines whether specific data is used during training. Mobile and edge deployments face elevated privacy risks when models trained on personal photos, health records, or behavioural patterns are stored locally and exposed to physical access or extraction attempts.

\textit{Integrity attacks} compromise the reliability and trustworthiness of on-device ML systems through model tampering, malicious manipulation of inference pipelines, and functional injection attacks. Unlike data extraction or model theft attacks, integrity attacks target the correct operation of deployed systems, resulting in compromised inference results and decreased model dependability. A prominent example is backdoor attacks, where an adversary can force the model to behave in a specific manner on targeted input, while the model maintains its normal behaviour on benign inputs. This type of malicious behaviour is especially concerning for mobile applications because they can often be downloaded from third party app stores and/or modified by a user after it has been deployed. 

The last category focuses on \textit{Availability attacks} which attempt to disrupt, restrict or completely prevent legitimate users from accessing a service. These attacks are typically carried out by overwhelming the system with requests that exhaust its computational resources. In the context of AI systems, this is usually achieved by crafting inputs or workloads that deliberately increase computational complexity.

\begin{figure}
  \makebox[\textwidth][l]{
    \hspace*{-1cm}
    \includegraphics[width=1.2\textwidth]{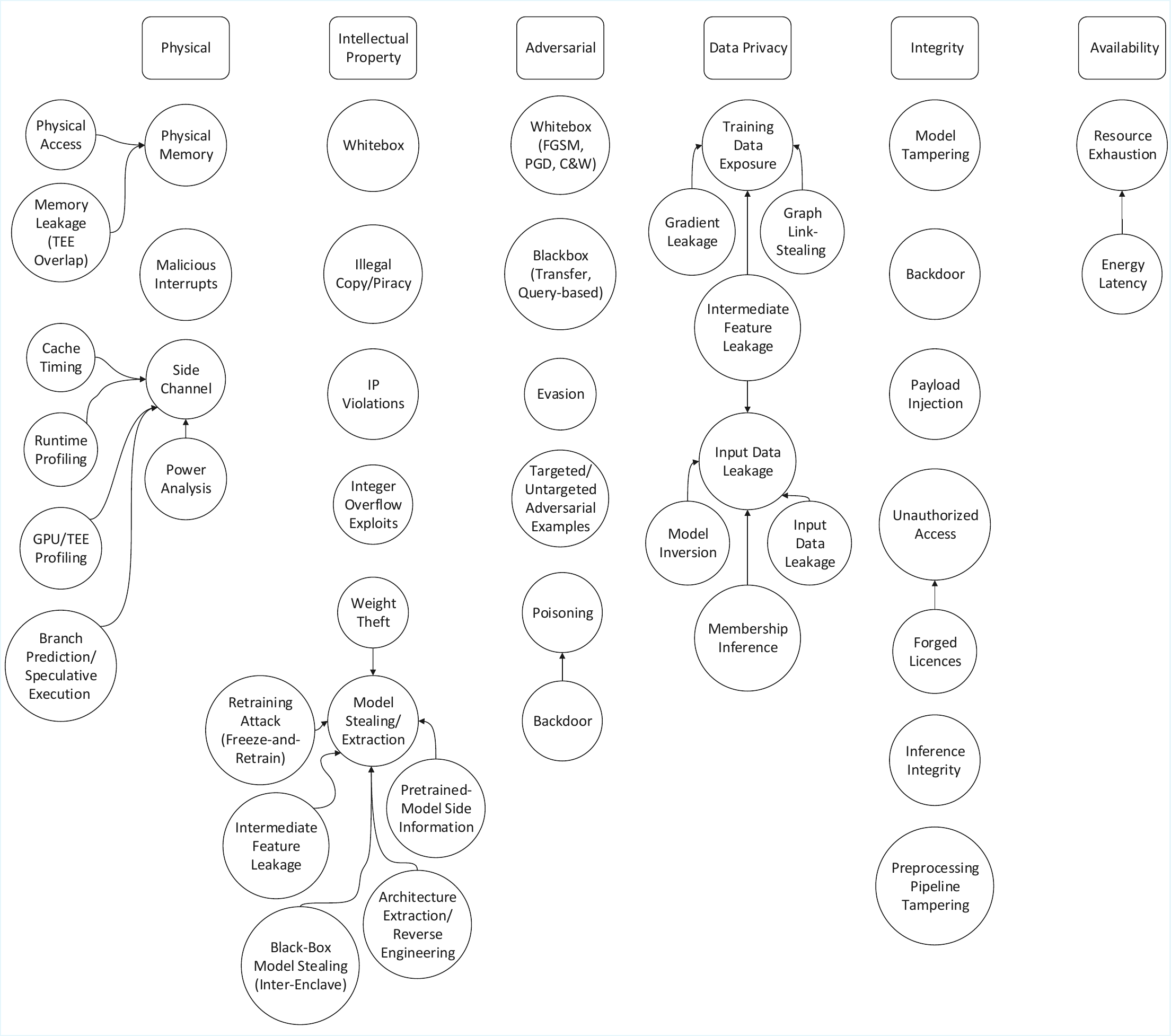}
  }
  
  \caption{Attack categorisation.}
  \label{fig:Attacks}
\end{figure}

Table~\ref{Tab:attacks} provides a high-level view of all major types of attacks, some examples of specific attacks (within each category), as well as their impact. Furthermore, Figure~\ref{fig:Attacks} provides a visualization of the main attack categories and their corresponding sub-categories, which are analysed in the attack and defence sections, where the relevant works are reviewed. The main categories are illustrated using rounded rectangles, while individual attacks are depicted as circles beneath each category. In some cases, specific attacks are further decomposed into sub-attacks, indicated by arrows pointing to their corresponding root attack. Additionally, Table~\ref{t:attack-effectiveness} gives an overview of the attack-related studies, in reverse chronological order per category. The effectiveness, target scale and attack category are also presented in the same table. In the following sections, we survey these attack studies and provide a brief description of each. Finally, Table~\ref{t:attack-summary} provides collective figures on the effectiveness of each attack category.

\begin{table}

\begin{center}
\begin{adjustwidth}{0cm}{}
\scriptsize
\renewcommand{\arraystretch}{1.5}
\begin{tabular}{|>{\centering\arraybackslash}m{2cm}|c|>{\centering\arraybackslash}m{2.5cm}|>{\centering\arraybackslash}m{2.5cm}|>{\centering\arraybackslash}m{3.5cm}
|} 
\hline
\textbf{Work} & \textbf{Year} & \textbf{Effectiveness} & \textbf{Target scale} & \textbf{Attack category} \\ 
\hline

\multicolumn{5}{|c|}{\textbf{Physical}} \\
\hline
\makecell{~RPMG\\~\cite{10738518}} & 2024 & 100\% (fixed opt.), 97.2\% (obfuscated) & TVM-compiled models, Arm Mali G52 & GPU/Accelerator Runtime Profiling \\
\hline
\makecell{PEDL\\~\cite{lee2022precise}} & 2022 & 5.8$\times$ improvement & Edge devices & Cache Side-Channel 
\\
\hline

\multicolumn{5}{|c|}{\textbf{Intellectual property}} \\
\hline
\makecell{SoK\\~\cite{298278}} & 2024 & 48.81\% all, 33.83\% unique & 210K Android apps & Weight Theft, Architecture Extraction \\
\hline
\makecell{DEMISTIFY\\~\cite{10.1145/3597503.3623325}} & 2024 & 82.73\% reuse & 1,511 mobile apps & Black-Box Model Stealing, Inter-Enclave \\
\hline
\makecell{REDLC\\~\cite{10771396}} & 2024 & 99\% recovery & 18 DL models & Architecture Extraction, Reverse Engineering \\
\hline
\makecell{REOM\\~\cite{10.1145/3597503.3639144}} & 2024 & 89.03\% vs 10.23\% & 244 TFLite models & Architecture Extraction, Reverse Engineering  \\
\hline
\multicolumn{5}{|c|}{\textbf{Adversarial}} \\
\hline
\makecell{TIM\\~\cite{10669107}} & 2024 & 47\% vulnerable & 1,250 apps & White-Box\\
\hline
\makecell{CYA\\~\cite{https://doi.org/10.1002/smr.2528}} & 2024 & 66.60\% success & 10 real-world apps & Black-Box \\
\hline
\makecell{GenDroid\\~\cite{XU2023103359}} & 2023 & 99.12\% (77 queries) & Query-Efficient & Black-Box \\
\hline
\makecell{SAA\\~\cite{9766323}} & 2022 & 71.7\% real-world & Android apps & Black-Box \\
\hline
\makecell{RODM\\~\cite{huang2021robustnessondevicemodelsadversarial}} & 2021 & 3$\times$ improvement & 10 applications & White-Box \\
\hline
\makecell{CYA\\~\cite{cao2021blackboxattacksdeeplearning}} & 2021 & Substitute Training & Behavioural Cloning & Black-Box \\
\hline

\multicolumn{5}{|c|}{\textbf{Data privacy}} \\
\hline
\makecell{Membership\\~\cite{chen2024private}} & 2024 & 92\% accuracy & 5 HAR datasets & Membership Inference\\
\hline
\makecell{Inversion\\~\cite{dong2021privacy}} & 2021 & PSNR: 18.83 & ResNet-50, 42 layers & Model Inversion\\
\hline

\multicolumn{5}{|c|}{\textbf{Integrity}} \\
\hline
\makecell{~Preprocessing\\~\cite{10.1145/3591197.3591308}} & 2023 & 81.56\% success & 320 MLKit apps & Inference Integrity\\

\hline
\makecell{~DeepPayload\\~\cite{9402020}} & 2021 & 93.5\% trigger detect & 54/116 apps & Tampering With Models\\
\hline

\multicolumn{5}{|c|}{\textbf{Availability}} \\
\hline

\makecell{~Energy\\~\cite{10.1145/3591197.3591307}} & 2023 & 25\% battery drain & Snapdragon 8Gen1 & Resource Exhaustion Attacks \\
\hline
\end{tabular}
\end{adjustwidth}
\end{center}

\caption{Quantitative analysis of attack effectiveness on on-device AI systems (PSNR: peak signal-to-noise
ratio).}
\label{t:attack-effectiveness}
\end{table}

\subsection{Physical Attacks}





Starting by analysing side-channel attacks, the work by \cite{lee2022precise} demonstrates that attackers can exploit shared cache memory on edge devices to extract model architecture and image dimensions through cache timing attacks such as Flush+Reload. This side-channel information improves model extraction effectiveness by up to 5.8 times compared to blind attacks, with image dimensions being the most critical information for successful extraction.

On the same front, the research in \cite{10738518} studies GPU runtime profiling attacks and shows that DNN architectures can be extracted from TVM runtime profiles on mobile GPUs by monitoring execution timing patterns. Using a binning technique that partitions operator durations, the attack achieves 100\% accuracy with fixed optimisation levels and 97.2\% accuracy even with obfuscated operator names, demonstrating that hardware timing side-channels persist despite software obfuscation efforts.

\subsection{Intellectual Property Attacks}



The widespread vulnerability of on-device models to direct extraction was comprehensively demonstrated by \cite{298278}, which analysed more than 200,000 Android applications and successfully extracted 16,500 ML models. Their framework categorises extraction attacks into four distinct approaches: app-based extraction targeting application packages (achieving 48.81\% success rate for all models and 33.83\% for unique models), device-based attacks exploiting runtime memory, communication-based attacks leveraging side-channel information, and model-based attacks using input-output query patterns. The authors have shown in this large scale study that even though advanced techniques for extracting data from memory and through side-channels can be used successfully in a testbed environment, they fail frequently when attempting to extract data from applications in a real-world environment due to the lack of reproducibility.

Building on direct model extraction, the automated tool DeMistify \cite{10.1145/3597503.3623325} escalates the attack not only by extracting models but also enabling their reuse through automated script generation. The tool first statically locates the ML models and then uses obfuscation-resilient program slicing to identify the minimum necessary execution components. It then automatically generates scripts to instrument mobile apps, a process that leverages boundary function recognition and dynamic instrumentation, which helps bypass restrictions like licence verification. DeMistify successfully reused $1,250$ out of $1,511$ top mobile apps in its evaluation, achieving a success rate of $82.73\%$. This empirical evaluation demonstrates that current model protection measures, including encryption and code obfuscation, are insufficient against such automated model reuse attacks.


Compiler optimisations introduce significant challenges to methods relying on reverse engineering techniques for model extraction, particularly for TVM-compiled models where layer fusion and operator obfuscation thwart traditional methods. REDLC \cite{10771396} addresses this by employing machine learning techniques—combining bidirectional LSTMs, MLPs, and graph convolutional networks to convert TVM-compiled binaries back to executable Keras models. This approach achieves 99\% accuracy in recovering 18 common deep learning models, subsequently enabling adversarial attacks that reduce target model accuracy by 88.65\% on average. This demonstrates that ML-based reconstruction techniques can achieve near-perfect recovery rates even against compiler-optimised models, representing a significant escalation in extraction threat sophistication.

Extending the reverse engineering threat, REOM~\cite{10.1145/3597503.3639144} addresses the issue of converting compiled TFLite models into debuggable PyTorch versions, enabling direct white-Box adversarial attacks. The framework successfully converted $92.6\%$ of 244 real-world TFLite models. White-Box attacks with the Projected Gradient Descent (PGD) method achieved a success rate of $89.03\%$ compared to just $10.23\%$ with regular Black-Box techniques. This proves that the assumed security of Black-Box attack techniques versus white-Box access coefficients was incorrect, vastly underestimating the threat that compiled on-device models pose; that is, model compilation is pointless if gradient access is granted.


\subsection{Adversarial Attacks}




Testing compiled inference-only models makes it impossible to evaluate their vulnerability. TIM \cite{10669107} uses automated code and model reconstruction to convert inference-only models into testable ones. Through this functionality, it has been determined that 47\% of commercial available applications tested by TIM have potential vulnerabilities to adversarial attacks, which highlights the security weaknesses in the models deployed.


Another direction in this category focuses on query-efficient Black-Box attacks. When full model access is unavailable, adversaries can exploit behavioural interfaces to mount effective attacks. The work in~\cite{https://doi.org/10.1002/smr.2528} demonstrates this by instrumenting DummyActivity components to collect input-output pairs and training substitute models that leverage adversarial transferability. This method achieves an average attack success rate of 66.60\%, across ten real-world applications, outperforming existing Black-Box methods by 27.63\%. Similarly, \cite{cao2021blackboxattacksdeeplearning} employs substitute model training techniques, demonstrating that even strong protection measures cannot prevent behavioural cloning attacks operating purely through model interfaces.

A critical challenge in mobile environments is minimising the number of query-based attacks to avoid detection. GenDroid \cite{XU2023103359} addresses this using genetic algorithms enhanced with Gaussian process regression and attention mechanisms, achieving a 99.12\% attack success rate against Drebin with only 77 queries on average, a significant improvement over existing methods that require over 1,000 queries. This query efficiency makes the attack substantially more practical and harder to detect in real-world deployments.

Mobile applications frequently employ transfer learning, fine-tuning pre-trained models for specific tasks. This widespread practice creates exploitable vulnerabilities. Smart App Attack \cite{9766323} implements a grey-box framework that extracts fine-tuned models, identifies their original pre-trained models, and constructs targeted adversarial examples, achieving success rates exceeding 70\% of real-world Android applications. Similarly, \cite{huang2021robustnessondevicemodelsadversarial} exploits the pre-trained model similarity by identifying the corresponding models from the TensorFlow Hub through similarity analysis, achieving success rates over 3 times higher than blind attacks across all 10 tested applications.

\subsection{Data Privacy Attacks}



Model inversion attacks attempt to reconstruct original training data from model parameters or intermediate features. The Divide-and-Conquer Inversion (DCI) method \cite{dong2021privacy} demonstrates that split computing---often assumed to provide privacy protection—remains vulnerable to sophisticated inversion attacks. This data-free approach can recover recognisable images from features after up to 42 convolutional layers by partitioning deep networks into shallow blocks and inverting each progressively using cycle-consistency techniques. The attack achieves high-fidelity reconstruction with Peak signal-to-noise ratio (PSNR) 18.83 and Learned Perceptual Image Patch Similarity (LPIPS) 0.443, without requiring real training data. This shows that even when models execute 35+ layers on-device before sharing features with cloud servers, attackers can still recover sensitive visual details.


In addition, Membership Inference Attacks (MIA) determine whether specific samples were included in a model's training dataset, potentially exposing sensitive information about individuals. Federated Human Activity Recognition, as shown by the authors in \cite{chen2024private}, demonstrates that these systems are susceptible to MIA. A malicious or curious server is able to classify the identity of a user's assigned client based on a sample of HAR data with up to 92\% accuracy by enumerating differences in user activity pattern. For the common defence techniques, such as L2 regularisation and dropout, their defences are weak, with the accuracy decreased by 10-20\%. This underscores severe privacy leakages of federated learning settings with wearable devices.




\subsection{Integrity Attacks}


Model tampering and malicious injection concern two significant issues, and some works have been focusing on a complete solution for those problems. DeepPayload \cite{9402020} serves as a case study in how researchers are targeting complete solutions for these twin challenges. In addition to demonstrating a White-Box backdoor attack on deployed deep-learning models via a neural payload injection, DeepPayload also demonstrated a Black-Box backdoor attack. While many traditional tampering methods rely on having access to the training data or requiring the model to be retrained in order to perform the tampering, DeepPayload is able to operate on the already-deployed and compiled versions of the models. Using reverse-engineering techniques, DeepPayload can inject malicious payloads into model data-flow graphs and provide malicious behaviour based on triggers that do not degrade the normal operation of the model. Additionally, DeepPayload was able to achieve successful trigger detection rates as high as 93.5\%, while operating at a very low level of computational overhead; therefore, it successfully compromised 54 out of 116 real-world Android applications. Therefore, this example clearly shows that even after models are deployed, compiled, and distributed, they are still vulnerable to post-deployment tampering and therefore threaten the integrity and trustworthiness of mobile ML systems.


Other works have been analysing Inference Pipeline Manipulation, such as the work by \cite{10.1145/3591197.3591308}, which demonstrates that attackers can target the data preprocessing pipeline rather than the models themselves, successfully compromising 81.56\% of 320 MLKit applications by injecting malicious code that modifies image parameters during preprocessing. This approach requires no model knowledge, highlighting that security considerations must extend beyond model protection to the entire ML inference pipeline.

\subsection{Availability Attacks}

Finally, some research has been introduced to combat Resource Exhaustion Attacks. The energy-latency attack \cite{10.1145/3591197.3591307} shows that mobile AI accelerators are vulnerable to availability attacks that deliberately reduce data sparsity in feature maps, achieving increased battery consumption while maintaining model accuracy on processors like Snapdragon 845 and Snapdragon 8Gen1; the battery consumption ranges from 14 to 21\%. The efficiency of modern mobile AI accelerators comes at a security cost: their reliance on sparsity-based optimisation makes them uniquely susceptible to specific attacks.

\section{Inference Protection Mechanisms}
\label{sec:defence}

During the last six years, a plethora of solutions have been proposed to mitigate the risks associated with on-device AI deployments. The defensive solutions included in this section aim to protect against assaults as those presented in Section~\ref{sec:attack}.
It should be noted that this analysis only considers research works that protect the inference data, the process, and the AI model itself, when the inference is executed on-device. On the contrary, works that employ AI to develop intrusion detection and prevention systems are excluded from our analysis. Furthermore, we only consider publications from 2019 to 2025, emphasising on the latest advances in this research area. 
In the rest of this section, we first develop a taxonomy for categorising the available countermeasures and then present the relevant publications.

\subsection{Taxonomy}
\label{subsec:taxonomy}

Depending on their focus, we classify the countermeasures into five primary categories, according to the main technology that assists in preventing an attacker from launching a successful attack. These categories are: model segmentation, access control, obfuscation, TEE optimisation and TEE extension. 
The countermeasures of each category can be used to prevent one or more of the attacks presented in Section~\ref{sec:attack}. It should be noted that each proposal may employ several mechanisms; however, this classification relies on the most pivotal one.

The first category, namely \textit{model segmentation}, is based on partitioning and decomposition to protect the model and user data~\cite{Deng_2020}. More specifically, the first strategy concerns the partitioning of the AI model with the aim of executing the critical layers inside a secure environment, and the rest in the unprotected section, namely Rich Execution Environment (REE). In this case, the processing of sensitive pieces of information is taking place in the secure area, that is, the TEE, and the results are communicated to the insecure host to complete the inference process. On the contrary, decomposition refers to the process of model factorisation/slicing into several units, often before or during the training phase. This approach aims at delivering an efficient and secure model that fits to the executing environment, overcoming resource constraints. Overall, model segmentation prevents adversaries from carrying out model stealing attacks and IP theft, MIA, model inversion, and the exploitation of data leaks.

The next category focuses on preventing the attacker from gaining access to the inference engine by enforcing \textit{access control} mechanisms. Many of the works in this category use authentication, licensing, and weight permutation to restrict access only to users that are aware of the underlying patterns. This type of countermeasure is used broadly in mechanisms that aim to protect against model and IP theft, unauthorised access and tampering, licence forgery and misuse, Key-Value (KV) leak, and cache attacks.

Model \textit{obfuscation} focuses on encoding certain pieces of information to conceal them from the adversaries, rendering reverse engineering assaults impractical. Existing works in the literature rely on two main techniques to obfuscate a model. The first considers the permutation of the model weights and parameters using secret or random patterns. The second type applies information reduction techniques to reduce the model accuracy. This is addressed by the works included in our analysis using model pruning and quantisation techniques. 
By employing this approach, the attacker only has access to the inputs and outputs of the model, rather than monitoring the inference engine internals. This category is effective in protecting the system from reverse engineering attacks, membership inference, model steal, and IP theft.

Finally, the last two categories relate to improvements in the TEE isolated region. Specifically, \textit{TEE optimisations} attempt to introduce enhancements to existing TEE architectures with the aim of executing more robust and accurate models. This is achieved with system-wide optimisations, such as efficient memory management, parallelisation strategies, and integration with accelerators. These improvements allow the execution of more computationally intensive AI models in the secure region, while increasing overall security. 
\textit{TEE extensions} introduce new features to extend the secure regions or their existing functionality to integrate with the latest architectures. Examples of this category introduce enhancements towards securing the computations performed in GPU memory, or deliver separate isolated environments for multi-model processing architectures.
Overall, TEE-related categories protect against data leaks and privacy leaks during inference.

\subsection{Model segmentation}

In this subsection, we analyse countermeasures that leverage partitioning and decomposition techniques. Table~\ref{t:defence-ai-tools-partitioning} offers an overview of each solution in reverse chronological order. The third column indicates which segmentation approach is followed by each proposal, i.e., partitioning vs decomposition. The fourth column presents the model family that is protected by the specific defence solution, while the following column summarises the attacks or threats addressed by each work. Finally, it presents a description of the defence technique that each solution proposes, as well as the type of device on which the AI model is deployed. 

\begin{table}
\renewcommand{\arraystretch}{1}  
\setlength{\extrarowheight}{2pt}
\scriptsize
\begin{tabular}{
|>{\centering\arraybackslash}p{1.5cm}
|p{0.7cm}| >{\centering\arraybackslash}p{0.6cm}| >{\centering\arraybackslash}p{0.7cm}|p{4cm}| p{3.7cm}| >{\centering\arraybackslash}p{1cm}|}
\hline
\textbf{Work} & \textbf{Year} & \textbf{P/D} & \textbf{AI} & \textbf{Attack} & \textbf{Countermeasure}&\textbf{Device} \\ 
\hline
\makecell{TSQP\\~\cite{defence2025}}& 2025&P& QNN & Model Stealing, Model Extraction, Integer Overflow, Inference Integrity & TSQP& Edge \\ 
\hline
\makecell{GNNVault\\~\cite{ding2025graphvaultprotectingedge}}& 2025&P &GNN & Model Stealing, Model Extraction, Graph Link-Stealing Attack & PBT Design with TEE-Secured Rectifier & Edge \\ 
\hline
\makecell{TBNet\\~\cite{liu2024tbnetneuralarchitecturaldefence}}&2024 & P&  DNN & Weight Theft & TEE-Based Two-Branch defence & Edge \\ 
\hline
\makecell{TEESlice\\~\cite{10.1145/3707453}}&2024&P/D &  \makecell{DNN,\\LLM} & Model Stealing, Model Extraction, Training Data Exposure, Membership Inference, Model Inversion  & PBT with TEE-Isolated Private Weights & Mobile \\ 
\hline
\makecell{Penetralium\\~\cite{YANG202430}}&2024&D &  DNN & Training Data Exposure, Membership Inference Attack, Model Inversion Attack &  Segmented TEE Inference with Confidence Perturbation & Edge  \\ 
\hline
\makecell{FakeNN\\~\cite{10831158}}& 2024&P& DNN & Model Stealing, Model Extraction & Segmented TEE Dual-Network with Backbone Degradation & Edge  \\ 
\hline
\makecell{MirrorNet\\~\cite{10323746}}& 2023 & P&DNN & Model Stealing, Model Extraction  & TEE-Protected Mirror Branch & Edge \\ 
\hline
\makecell{LEAP\\~\cite{9895274}}& 2023 & P&DL &    Model Stealing, Model Extraction, Membership Inference & Lightweight Parallel TEE & Mobile\\ 
\hline
\makecell{SecureQNN\\~\cite{10.1007/978-981-99-7969-1_1}}& 2023 & P&QNN & Weight Theft, Freeze-and-Retrain & QNN Layer Segmentation in TrustZone-M & \makecell{IoT} \\ 
\hline
\makecell{TEESlice\\~\cite{zhang2023privacyleftoutsideinsecurity}}& 2023 & P/D& DNN & Weight Theft, Pretrained-Model Side Information Attack, Freeze-and-Retrain, Membership Inference Attack, Model Inversion Attack & PBT with TEE-Isolated Privacy Slices & \makecell{Edge} \\ 
\hline
\makecell{T-Slices\\~\cite{10.1145/3577923.3583648}}&2023&D&  \makecell{DL,\\CNN} & Weight Theft, Intermediate Feature Leakage & Dynamic Layer Slicing in TrustZone & IoT\\ 
\hline
\makecell{ShadowNet\\~\cite{sun2023shadownetsecureefficientondevice}}& 2023 &P& ML & Weight Theft, Architecture Extraction, Reverse Engineering, Intermediate Feature Leakage & Secure GPU Offload via Weight Transformation & \makecell{Mobile}\\ 
\hline
\makecell{Hybridtee\\~\cite{9358260}}& 2020 & P&DNN & Weight Theft, Architecture Extraction, Reverse Engineering, Intermediate Feature Leakage, Inference Integrity  & SRAM-Optimized Secure Inference with Encrypted DRAM & Mobile \\ 
\hline
\makecell{DarkneTZ\\~\cite{Mo_2020}}& 2020 & P& DNN & Membership Inference Attacks, Intermediate Feature Leakage  & Hardware-Assisted Layer Partitioning for Privacy-Preserving DNNs & Edge \\ 
\hline
~\cite{vannostrand2019confidentialdeeplearningexecuting} & 2019 & P&DNN  & Weight Theft, Architecture Extraction & TEE-Based Confidential Inference for Proprietary Deep Learning Models & Mobile\\ 
\hline
\end{tabular}
\caption{Countermeasure proposals based on model segmentation (P: partitioning, D: decomposition).}
\label{t:defence-ai-tools-partitioning}
\end{table}

Based on the table, it seems that most of the solutions apply partitioning, except two solutions that make use of decomposition and one combining both techniques. The majority of the research investigates countermeasures on DNN deployments, while four distinct works focus, respectively, on LLM, Graph Neural Network (GNN), Convolutional Neural Network (CNN) and Quantisation Neural Networks (QNN). As far as it concerns the attacks, aggressors rely mostly on model stealing/extraction and weight theft pattern to exploit the targeted models. These categories dominate with seven and six works respectively. It is worth noting that two distinct researches introduce the integer overflow vulnerability and the graph link-stealing attack. Regarding the proposed countermeasures, the main focus is placed on the elimination of the attack surface by hiding weights, activations and graph edges. A more detailed view of each research is given bellow.

The work in~\cite{defence2025} investigates model stealing attacks on QNN executed in edge environments. The authors introduce a partitioning solution, namely TEE-Shielded QNN Partition (TSQP), that transforms White-Box inference to Black-Box, by isolating certain scales in TEE. The scales denote the compression factor of a float so that it can fit into the integer range. In this respect, by shielding the scales in TEE, the attacker has access only to the quantised integer result that is conveyed in REE. On the same direction, the proposal provides a solution to the 8-bit quantisation issue that results in severe integer overflow vulnerabilities during inference~\cite{DBLP:journals/corr/abs-1712-05877}. To combat these issues, TSQP introduces parameter de-similarity (PDS) techniques which are effective against model stealing attacks, while in parallel they safeguard inference integrity. 

The authors of~\cite{ding2025graphvaultprotectingedge} present GNNVault, a secure method to protect the privacy and intellectual property of the model, building over the partition-before-training (PBT) strategy. Using a hybrid approach, the authors deploy two different models, namely the backbone and the rectifier, trained on public and private data, respectively. The first model is deployed in REE, while the second, i.e., the rectifier, in TEE. This design aims at producing rough embeddings in the REE, which are being rectified from the secure model that uses the real private graph in the secure enclave. 

The work in~\cite{liu2024tbnetneuralarchitecturaldefence} introduces TBNet, a two-branch solution to safeguard DNN models from stealing attacks. This partitioning solution relies on two identical models running in the REE and TEE regions, respectively. 
During the training phase, the insecure model runs in the insecure area (REE) for performance, while being iteratively pruned so that it becomes useless to a potential attacker. In parallel, it transfers knowledge to the secure model residing in the secure region (TEE).
At the inference step, the outputs from the two branches are merged inside the TEE and the result is filtered from a last classification layer which gives the prediction result. The only piece of information that is communicated outside of TEE is the final result.

In the same vein, the research in~\cite{YANG202430} introduces penetralium, a system to protect model privacy on edge environments. The proposed solution protects against model steal attacks, membership inference attacks, and model inversion attacks. The architecture leverages a computational engine (SGX engine) and a model converter that gradually decomposes the model into segments that can fit the TEE limitations. The sensitive pieces are executed in the secure region, while the rest in REE. After inference is completed, penetralium applies a lightweight perturbation policy to the model probability scores, so that the attacker cannot exploit them to launch privacy inference attacks. 

The authors in~\cite{10831158} introduce FakeNN, a secure inference framework, to combat model steal and adversarial attacks. FakeNN relies on a dual-architecture that splits and stores the model in a backbone and a lightweight network, respectively. One of the main advantages of the introduced architecture pertains to the limitation of the learning ability of the backbone model. Specifically, this aims at reducing the learning ability of the model, so that the only accurate inference result is obtained from the lightweight network inside the secure area. Finally, the authors enhance even more the secure network performance by integrating a channel attention mechanism that focuses on the important information channels.

The work in~\cite{10323746} presents MirrorNet, a dual-layer partitioning framework to protect DNN deployments from model stealing attacks on edge devices. MirrorNet creates a lighter version of the DNN model that can be executed inside the TEE with minimal delays. It operates with two main components, the backbone model and the companion partial monitor (CPM) running inside the TEE. The last one resides in the secure region and corrects the intermediate layer results of the BackboneNet, ensuring model confidentiality. Every time, during inference, CPM corrects the intermediate outputs generated from the BackboneNet using the secret knowledge stored in the TEE.

The authors in~\cite{9895274} present LEAP, a partitioning method to protect against model inference attacks in mobile devices. LEAP is a lightweight developer-friendly TEE, devoted to mobile apps, that is used as an alternative to ARM TrustZone. This lightweight system enables the execution of parallel protection sandboxes with easy access to peripherals like GPU, thus improving system flexibility. This setting enables the execution of multiple models, or intelligent applications to be executed securely within the same environment, thus making LEAP a very good candidate for multi-model execution on smartphones.

The work in~\cite{10.1007/978-981-99-7969-1_1} explores the use of TrustZone-M to protect QNN privacy. The authors introduce SecureQNN, a partitioning framework that assesses the attacker's effort on building a model with the same accuracy. SecureQNN traverses the model layer by layer and evaluates which layers are more sensitive. That is, it selects those that require the minimum training effort from the attacker's perspective. Those layers are loaded into the Trustzone-M area, while the rest remain outside, in the insecure regions.

The authors in~\cite{10.1145/3707453} conduct a benchmark of several TEE-shielded DNN Partition (TSDP) solutions and highlight their limitations. The authors extend their own previous work by introducing TEESlice, a novel hybrid strategy to protect sensitive neural network models in TEE~\cite{zhang2023privacyleftoutsideinsecurity, 10.1145/3536168.3543299}. A significant security issue addressed in these works, concern model stealing and membership inference attacks, where an attacker may compromise the model weights and user data at the TEE layer. To address this issue, TEESlice implements a combination of partitioning and decomposition techniques, specifically a partition-before-training approach, which effectively separates privacy-sensitive weights from other components of the model and provides Black-Box level protection by only shielding slices in TEEs. The proposed solution is designed for TEE-based systems, such as edge and mobile devices. 

The research in~\cite{10.1145/3577923.3583648} proposes T-Slices, a novel system to protect the integrity and confidentiality of DL models on IoT devices. The proposed solution overcomes the resource footprint limitations of executing DL models in the ARM TrustZone by decomposing each model layer into smaller slices. During execution, each of the slices is sequentially loaded into TrustZone and the decryption takes place gradually. Bear in mind that batch processing is also possible by sending multiple layers for processing at once. The output of each layer is kept securely in the secure region, so that each new layer can process the previous output. In contrast to other solutions, the proposed system does not modify the original model, and thus the original prediction accuracy remains unaffected.

The paper in~\cite{sun2023shadownetsecureefficientondevice} presents Shadownet, a partitioning architecture to protect against model stealing attacks. The solution focuses on the problem of TEE-based solutions which operate inside the secure area and do not take advantage of the AI accelerators running in REE. More specifically, the proposed approach outsources the computationally heavy linear model layers in the untrusted accelerators. This is achieved by transforming the weights in a way that cannot be recovered by an attacker. After the computations are completed, the results are securely reverted again inside the TEE.

The authors of~\cite{9358260} proposed Hybridtee, a system that protects DL and LLMs running on mobile devices against threats, such as model stealing and data privacy leakage at the TEE layer. Specifically, Hybridtee implements a novel partitioning approach that strategically divides the DL model into local and remote components, executing the sensitive parts in a local TEE and the less sensitive parts in a remote TEE. The proposed solution is designed for devices that support ARM TrustZone and Intel SGX, such as smartphones and edge servers. Hybridtee also ensures the security and integrity of the model inference process by using secure communication protocols and encryption techniques to protect the data exchanged between local and remote TEEs. The authors evaluated Hybridtee on a range of DL models, including Darknet19 and GoogLeNet, and reported promising results in terms of security, performance, and scalability.

The authors of~\cite{Mo_2020} propose DarkneTZ, a framework that protects DL models, including LLMs, on edge devices. A significant security issue addressed in the paper is model stealing and membership inference attacks, to steal the model weights and user data at the TEE layer. DarkneTZ implements a partitioning strategy that separates sensitive layers of the model and executes them in a trusted environment, ensuring the security and integrity of the model inference process. The proposed solution is designed for devices that support ARM TrustZone, such as smartphones and IoT devices.

The authors of~\cite{vannostrand2019confidentialdeeplearningexecuting} present a partitioning system that protects DL and LLMs running on ARM-based devices, such as smartphones and IoT devices, by leveraging the TrustZone security extension. The proposed system implements a secure inference framework that uses the TrustZone environment to protect the model and user data, ensuring the security and integrity of the model inference process. The system has been evaluated on a range of DL models, including MobileNet and ResNet, and has shown promising results in terms of security, performance, and power consumption.

\subsection{Access control}

This part offers an analysis of works that employ access control mechanisms to prevent unauthorised access to the model internals and the inference process. Table~\ref{t:access-control-table} provides an overview of the access control solutions examined in this section. 

\begin{table}
\renewcommand{\arraystretch}{1}
\scriptsize
\setlength{\extrarowheight}{2pt}
\begin{tabular}{
|>{\centering\arraybackslash}p{2.8cm}
|c
|>{\centering\arraybackslash}p{1cm}
| p{2.3cm}
|p{2.5cm}
| >{\centering\arraybackslash}p{1.5cm}|}
\hline
\textbf{Work} & \textbf{Year} &\textbf{AI} & \textbf{Attack}    &\textbf{Countermeasure}&\textbf{Device} \\ 
\hline
~\cite{10978383}& 2025& CNN & Black-Box Model Stealing, Forged Licenses   & Style-Driven Licensing & Edge \\ 
\hline
~\cite{fi17020085}& 2025 & TinyML & Tampering With Models, Unauthorized Access    & EAT-Based Dual Attestation for TinyML and Platform Integrity & Edge \\ 
\hline
\makecell{IoTCloak\\~\cite{10.1145/3722566.3727630}}& 2025 & TinyML & Tampering With Models, Unauthorized Access   & Cortex-M Secure Inference via TEE and Watchpoints & IoT \\ 
\hline
\makecell{CoreGuard\\~\cite{li2024coreguardsafeguardingfoundationalcapabilities}}& 2024& LLM & Black-Box Model Stealing  & TEE-Protected Row/Column Permutation for LLM Anti-Stealing & Edge \\ 
\hline
\makecell{EdgePro\\~\cite{10433683}}&2024 &  DL & Black-Box Model Stealing, Unauthorized Access  & Neuron-Level Password-Based Model Lock & Edge\\ 
\hline
\makecell{TransLinkGuard\\~\cite{Li_2024}}& 2024 & LLM &Black-Box Model Stealing, Unauthorized Access  & TEE-Protected Row/Column Permutation Locking & Edge \\ 
\hline
\makecell{SecureDL\\~\cite{272264}} &2021 & DNN & Model Stealing, Model Extraction  & N/A & Mobile \\ 
\hline
\makecell{TCFDL\\~\cite{ZHANG2021108055}}& 2021 & ML & Input Data Leakage, Inference Integrity   & TEE-enablead trusted collaboration framework & IoT \\ 
\hline
\end{tabular}
\caption{Countermeasure proposals based on Access Control.}
\label{t:access-control-table}
\end{table}

The research in~\cite{10978383} presents a solution to protect against intellectual property hijacking and unauthorised access to AI models running on edge devices. The proposed mechanism leverages the image style aspect to create a licensing mechanism for access control. That is, the model is trained using certain image style data, used as a licence, so that any inappropriate input with misaligned data will make the model fail.
The process is driven by the licence generator, which consists of a lightweight, real-time style transfer model that is executed directly on the edge. The aim of this component is to embed a predefined style into the models operational framework, so that a potential attacker is prevented from executing successful inference requests.  

The research in~\cite{fi17020085} introduces a dual attestation mechanism to protect from tampering and unauthorised access assaults against TinyML models running in IoT devices. This work bridges a literature gap by ensuring both the integrity of the AI model and the device. The core of the proposal relies on Entity Attestation Tokens (EAT). More specifically, the authors propose the creation of the first token inside the TEE that signs claims related to the device. The second token, namely Machine Learning Entity Attestation (ML-EAT), contains details about the model, and it is signed with keys provisioned securely into the secure region.

The authors of~\cite{10.1145/3722566.3727630} propose IoTCloak to protect IoT systems from model integrity attacks. The system leverages hardware watchpoints to ensure that specific regions of the memory used to store the ML code and models cannot be altered. Additionally, the authors introduce cryptographic accelerators (CRYA) to compute SHA-256 digests for sensitive pieces of information that watchpoints cannot cover. Finally, an additional feature of this work is the handling of watchpoint‑triggered write exceptions, which are routed to the secure sections so that they cannot be tampered with.

The research in~\cite{li2024coreguardsafeguardingfoundationalcapabilities} presents CoreGuard, a mechanism to prevent LLM steal in edge devices. This work focuses on the protection of general-purpose LLMs, and their foundational capabilities, which differentiate from existing solutions that usually monitor task-related parameters. This is why the authors introduce the term foundational capability stealing, which indicates that a stolen edge-deployed LLM can be used as a backbone for generic tasks, leveraging the general reasoning ability of the model. Regarding the proposed protection, in the first phase the system relies on model locking, a process that permutes the weights of the linear layers, to restrict access by rendering the model non-functional. The user needs to know how to swap the input so that they match the permutation pattern, and the authorisation is successful. In the second phase, the authors introduce an efficient self-propagation mechanism to automatically permute the outputs of each layer, saving this way communication overhead. 

The work in~\cite{10433683} presents EdgePro, a solution to prevent model steal and unauthorised model access on edge devices. Instead of encrypting the complete model, EdgePro selects and locks certain neurons which are used as authorisation points. To achieve that, the proposed solution alters the activation values of the selected neurons, with the aim of preventing attackers from using the model if they are not aware of the pattern. During inference, model activations are compared with stored configuration values to identify that the model is executed in the legitimate environment. 

The authors of~\cite{Li_2024} propose TransLinkGuard, a plug-and-play transformer model protection approach, which can protect LLMs and other transformer-based models against model stealing attacks. A significant security issue addressed in the paper is model stealing in the TEE layer. To counter this issue, TransLinkGuard implements a permutation strategy to protect the model weights, and integrates the authorisation mechanism with a linear layer, which involves more parameters and makes the authorisation process harder to violate. The proposed solution is designed for devices that support TEEs, such as edge devices. 

The authors of~\cite{272264} propose SecureDL, a system that aims to protect DL models, including LLMs, on edge devices from model stealing. To achieve this, SecureDL implements a secure inference framework that uses homomorphic encryption and differential privacy to protect model and user data. The proposed solution is designed for devices that support ARM TrustZone, such as smartphones and IoT devices.

The authors in~\cite{ZHANG2021108055} introduce a trusted and collaborative framework for deep learning (TCFDL) for IoT and edge devices. The paper addresses the limitations of IoT devices in executing AI models on edge nodes, while considering privacy implications. The framework relies on a hardware-based TEE module and introduces two different mechanisms to ensure data transmission security. 
The first one uses traditional TLS, but each intermediate proxy performs decryption and re‑encryption inside a TEE. In the second, the data remain encrypted throughout transit since in each proxy they are re-encrypted without previously being decrypted.

\subsection{Obfuscation}

This subsection analyses works that employ obfuscation schemes to prevent privacy leakages. An overview of the obfuscation-based defence mechanisms discussed in this section is summarised in Table~\ref{t:obfuscation-table}.

\begin{table}
\renewcommand{\arraystretch}{1}
\setlength{\extrarowheight}{2pt}
\scriptsize
\begin{tabular}{|>{\centering\arraybackslash}p{2.2cm}|
>{\centering\arraybackslash}p{0.7cm}|
>{\centering\arraybackslash}p{0.7cm}|
p{3.1cm}|
p{3cm}|
>{\centering\arraybackslash}p{1cm}|
}
\hline
\textbf{Work} & \textbf{Year} &\textbf{AI} & \textbf{Attack}   &\textbf{Countermeasure}&\textbf{Device} \\ 
\hline
\makecell{GroupCover\\~\cite{zhang2024groupcover}}& 2024& DNN & Weight Theft, Black-Box Model Stealing, Pretrained-Model Side Information, Architecture Extraction & TEE-Based Randomized Model Obfuscation & Edge, IoT  \\ 
\hline
~\cite{10.1145/3665314.3670821}&2024 &  DNN & Membership Inference, Data Leakage during Inference & Layer-Isolation Obfuscation Against MIAs & Edge \\ 
\hline
\makecell{KV-Shield\\~\cite{10.1145/3691555.3696827}}& 2024 & LLM & Intermediate Feature Leakage   & TEE-Secured KV Permutation Obfuscation & Mobile, IoT \\ 
\hline
\makecell{CustomDLCoder\\~\cite{zhou2024modellessbestmodelgenerating}}& 2024 & DL & Model Stealing, Model Extraction  & Model-Less Deployment for Model Hiding & Mobile, IoT \\ 
\hline
\makecell{Dynamo\\~\cite{Zhou_2024}}& 2024 & DL & Architecture Extraction, Reverse Engineering  & Homomorphic-Inspired Runtime Obfuscation & Mobile, IoT  \\ 
\hline
\makecell{Modelobfuscator\\~\cite{10.1145/3597926.3598113}}&2023 & DL & Architecture Extraction, Reverse Engineering & Reverse-Engineering–Resistant Model Obfuscation & Mobile, IoT \\
\hline
\makecell{Model Protection\\~\cite{9609559}}& 2022 & ML & Weight Theft  & Enclave-Controlled Model Degradation defence & Edge \\ 
\hline
\makecell{MMGuard\\~\cite{9474328}}& 2021 & DNN & Illegal Copy/Piracy, Tampering with models   & App-Anchored Model Obfuscation & Mobile \\ 
\hline
\end{tabular}
\caption{Countermeasure proposals based on Obfuscation techniques.}
\label{t:obfuscation-table}
\end{table}

The research in~\cite{zhang2024groupcover} presents GroupCover, a scheme that protects DNN models from stealing attacks. The proposed solution prevents attackers from exploiting the inherent randomness vulnerability of existing model‑obfuscation techniques. The solution operates in two phases, the first one performed offline is devoted to preprocessing and model obfuscation, whereas the second one is online and concerns inference. During the offline step, the private model parameters are randomly obfuscated, without following a specific strategy. 
Because the resulting obfuscation is indistinguishable from random guessing, an attacker gains no useful information, making the protection robust against inference attacks.
In the online inference step, the obfuscated model is used for the computations in the GPU, and the relevant results are reconstructed inside the TEE using some random parameters to safely reverse the result. 

The authors in~\cite{10.1145/3665314.3670821} present a solution to protect edge AI deployments from membership inference attacks. In the initial step, the partitioning technique is applied to split the model into two parts: the first is stored in the REE, while the second one, which contains the more sensitive layers, is executed inside the TEE.
By protecting a configurable number of layers inside the TEE, the attacker cannot access their output and thus the MIA attack is impractical. On top of that, the authors apply quantisation in order to optimise the inference process and reduce the memory footprint. This obfuscation effect results in noisy outputs that cannot be reverse-engineered by an aggressor. 

The authors of~\cite{10.1145/3691555.3696827} investigate LLMs deployed on edge devices equipped with GPUs. They consider protection against KV-pair leakages emitted during the inference process. These pieces of information can be eavesdropped by an attacker to reconstruct the complete conversation, and thus expose sensitive data. To protect against this vulnerability, the authors introduce KV-Shield, a system that protects KV pairs using weight permutation. That is, during the initialisation phase, the weight matrices are permuted, and thus the KV pairs are scrambled. During the runtime phase, the KV pairs used by the attention mechanism are restored to their original form inside the TEE, so they are never exposed to potential attackers.

The authors of~\cite{zhou2024modellessbestmodelgenerating} propose CustomDLCoder, a novel method to generate pure code implementations of DL models, which can protect DL models against model stealing and reverse engineering attacks. 
This security threat is particularly relevant at the on-device deployment layer. To counter this issue, CustomDLCoder implements a code generation approach that removes explicit model representations, making it difficult for attackers to extract model information. The proposed solution is designed for devices such as smartphones, including Android devices, and can be used to deploy various DL models, including MobileNet and GPT-2.

The authors of~\cite{Zhou_2024} propose Dynamo, a system that protects DL models on mobile devices. A substantial threat addressed in the paper is model stealing, to leak the model weights and user data. To counter this issue, Dynamo implements a dynamic model obfuscation strategy that couples obfuscated DL model operators, which ensures the security and integrity of the model inference process. The proposed solution is designed for devices such as Android smartphones.

Research~\cite{10.1145/3597926.3598113} proposes a novel system called Modelobfuscator which aims to protect DL models on mobile devices by obfuscating the model's structure and parameters. The proposed solution overcomes the model stealing issue by making it harder for attackers to extract the model's information, without affecting the original model's prediction accuracy or modifying the model itself. This is achieved through five obfuscation strategies, including renaming, parameter encapsulation, and neural structure obfuscation, which effectively resist model parsing tools and provide extra security for smart apps, despite the fact that this comes with the cost of an increased deliverable size.

The authors of~\cite{9609559} propose Model Protection, a system that protects DL models, including CNNs, running on edge devices. A significant security issue addressed in the paper is model stealing, where an attacker may steal the model weights and user data at the TEE layer. To address this issue, the authors propose a secure enclave-based inference service that adds crafted random values to the model weights, making it difficult for attackers to steal the model. The proposed solution is designed for devices that support Intel SGX, such as edge servers and smartphones.

The authors of~\cite{9474328} propose MMGuard, a system that protects DL models, including LLMs, on Android devices. A significant security issue addressed is model stealing and tampering, where an attacker may leak the model weights and the user data. To counter this issue, MMGuard implements a mutual authentication mechanism between the Android app and the DL model, which ensures the security and integrity of the model inference process. In order to achieve that, the authors introduce a new input branch that takes owner-related signature information to enforce access control. The proposed solution is designed for devices such as Android smartphones.

\subsection{TEE optimisations}

This subsection investigates works that optimise the TEE environment to reduce resource footprint while strengthening the security guarantees of on-device model execution.
The works analysed in this category are summarised in Table~\ref{t:tee-optimizations}.

\begin{table}
\renewcommand{\arraystretch}{0.8}
\scriptsize
\renewcommand{\arraystretch}{1}
\setlength{\extrarowheight}{2pt}
\begin{tabular}{|
>{\centering\arraybackslash}p{1.5cm}|
c|
>{\centering\arraybackslash}p{1cm}|
p{3cm}|
p{3.5cm}|
>{\centering\arraybackslash}p{1.5cm}|
}
\hline
\textbf{Work} & \textbf{Year} &\textbf{AI} & \textbf{Attack} &\textbf{Countermeasure}&\textbf{Device} \\
\hline
\makecell{Smartzone\\~\cite{10949698}}& 2025& DNN, LLM & Model Stealing, Tampering/Fault Injection, Data Leakage During Inference & Multi-Optimized Trusted Inference on TrustZone & IoT \\ 
\hline
\makecell{Smart-Zone\\~\cite{xie2024memoryefficientsecurednninference}}& 2024& DNN & Data Leakage During Inference & Adaptive Memory-Optimized TEE Inference Runtime & IoT  \\ 
\hline
~\cite{10.1145/3579856.3582820}& 2023 & DNN & Membership Inference Attack, Inference Integrity & TEE-Optimized Pruning and Partitioning & Mobile\\ 
\hline
\makecell{GuardiaNN\\~\cite{10.1145/3528535.3531513}}& 2022& DNN & Physical Memory Attacks & TEE-Accelerated Secure DNN Execution Through Memory and Crypto Offloading & Mobile, IoT\\ 
\hline
\makecell{SecDeep\\~\cite{10.1145/3450268.3453524}}& 2021 & DNN & Data leakage during inference, Inference Integrity & Secure GPU-Accelerated Inference Optimization in TrustZone & Mobile, IoT\\ 
\hline
\makecell{OMG\\~\cite{Bayerl_2020}}& 2020 & RNN & Data leakage during inference & TEE-Based Secure Offline Inference Isolation in ARM TrustZone & Mobile \\
\hline
\end{tabular}
\caption{Countermeasure proposals based on TEE optimisations.}
\label{t:tee-optimizations}
\end{table}

The authors of~\cite{10949698} propose Smartzone, a runtime support for secure and efficient on-device inference on ARM TrustZone, which can protect various AI models, including DL models and LLMs. A significant security issue addressed in the paper is model stealing and data leak at the TEE layer, where the model inference takes place. To counter this issue, Smartzone implements a trusted inference-oriented operator set, proactive multi-threading parallel support, and on-demand secure memory management, ensuring the security and integrity of the model inference process. The proposed solution is designed for devices that support ARM TrustZone.

The authors in~\cite{xie2024memoryefficientsecurednninference} present Smart-Zone, a novel memory management solution that aims to enhance model and inference data privacy. The approach relies on the adaptation of the secure memory size to fit the demands of the pre-trained DNN models. This is achieved by adapting the memory priorities based on the model being used. Moreover, it assigns priorities to memory regions in order to avoid conflicts. Also, the authors present Tinylib, a lightweight library designed for efficient inference inside the TrustZone. This approach does not target model partitioning; rather, it leverages both the secure and non secure worlds to complete the inference.

In~\cite{10.1145/3579856.3582820}, the authors acknowledge the problem of executing DNN models on mobile platforms and present a novel approach to securely execute such models on mobile devices. 
First, they propose to prune the model by removing redundant neurons that do not affect the overall accuracy. By using this approach, the authors optimise model execution as they preserve only the necessary parts. The model is also partitioned to fit and execute efficiently in the mobile TEE. After applying the pruning strategy, the proposed solution deallocates the occupied memory to improve the overall footprint.

The authors of~\cite{10.1145/3528535.3531513} present GuardiaNN, a system to tackle physical attacks, such as cold boot, on mobile and embedded devices. The authors focus on the DRAM region of a TEE which stores unencrypted sensitive data. That is, secure regions such as TrustZone do not offer protection on the memory level, and thus they become easily exploitable by aggressors. The proposed solution relies on the encryption of sensitive data in DRAM so that they remain protected, but it loads and decrypts the data in SRAM when required. To reduce the swap overhead, the authors propose to use direct convolutions, which reduces DRAM accesses. Additionally, they propose mechanisms to re-use the data in SRAM, so that swapping is eliminated. Finally, the CPU is decoupled from the encryption/decryption overhead and these operations are executed in dedicated hardware inside the SoC.

The authors of~\cite{10.1145/3450268.3453524} propose SecDeep, a system that protects DL models, including LLMs, against model stealing and data privacy leakage issues, where an attacker may compromise the model weights and user data at the TEE layer. To counter this issue, the authors implemented a secure inference framework that uses a combination of hardware-based TEE and software-based encryption techniques to protect the model and user data. The proposed solution is designed for devices that support ARM TrustZone, such as smartphones and IoT devices. SecDeep also ensures the integrity of the model and the underlying computation framework, and can maintain the performance of DL inference on the edge by securely interfacing on-device accelerators with TEEs. Additionally, SecDeep has been evaluated in a range of DL models, including SqueezeNet and MobileNet, and has shown promising results in terms of security and performance.

The authors in~\cite{Bayerl_2020} introduce the Offline Model Guard (OMG). The solution focuses on securing the data, model and processing algorithms using an optimised TEE-based architecture. The solution targets ARM devices and leverages TrustZone to achieve isolation among the different hardware components. More specifically, this work aims to overcome the performance challenges of traditional TEE solutions, by leveraging unprivileged user-space enclaves using SANCTUARY~\cite{Brasser2019SANCTUARYAT}. That is, OMG takes advantage of the performance of the insecure region that is secured using TrustZone components like the TrustZone Address Space Controller (TZASC) and the TrustZone Protection Controller (TZPC).

\subsection{TEE extensions}

This part presents the works that extend TEE capabilities, enhancing security against data leaks and privacy breaches. More specifically, it includes solutions that make use of the GPU or investigate multi-enclave deployments.
Table~\ref{t:tee-extensions} offers an overview of the solutions in reverse chronological order.

\begin{table}

\renewcommand{\arraystretch}{1}
\setlength{\extrarowheight}{2pt}
\scriptsize
\begin{tabular}{|
>{\centering\arraybackslash}p{1.5cm}|
c|
>{\centering\arraybackslash}p{1cm}|
p{3.5cm}|
p{2.5cm}|
>{\centering\arraybackslash}p{1.5cm}|
}
\hline
\textbf{Work} & \textbf{Year} &\textbf{AI} & \textbf{Attack} &\textbf{Countermeasure}&\textbf{Device  Type} \\ 
\hline
~\cite{abdollahi2025earlyexperienceconfidentialcomputing}& 2025 & ML & Model Stealing, Model Extraction, Membership Inference, Memory Leakage, Tampering with models & Arm CCA TEE extension for Low-Overhead Confidential Inference & Mobile, IoT\\ 
\hline
\makecell{ASGARD\\~\cite{moonasgard}}& 2025& DNN & Model Stealing, Model Extraction, Data Leakage During Inference, Unauthorized Access, Tampering with Models, Memory Leakage, Side Channels & Virtualization-Extended TEE Boundary for Secure Accelerated Inference & Mobile \\ 
\hline
~\cite{10637594}& 2024 & ML & Black-Box Model Stealing (Inter-Enclave), Input Data Leakage (Malicious Model) & Multi-Enclave TEE Extension for On-Device ML Isolation & Mobile, IoT\\ 
\hline
\makecell{Devlore\\~\cite{bertschi2024devloreextendingarmcca}}& 2024 & N/A & Malicious Interrupts, Memory Leakage & TEE-Extended Peripheral Access via CCA Realm Isolation & Mobile, IoT \\ 
\hline
\makecell{GuaranTEE\\~\cite{10.1145/3642970.3655845}}& 2024 & ML & Model Stealing, Model Extraction, Tampering with Models & CCA Realm Enclaves for Model Privacy and Verifiable Edge Inference & Mobile, IoT\\ 
\hline
\makecell{StrongBox\\~\cite{10.1145/3548606.3560627}}& 2022& NN & Memory Leakage, Unauthorized Access & TrustZone-Extended GPU TEE for Secure Unified-Memory Acceleration & Mobile, IoT \\ 
\hline
~\cite{RAHMAN2020103737}& 2020 & ML & Training Data Exposure & TEE-Boundary Extensions with Homomorphic QoS Protection & Edge\\
\hline
\end{tabular}
\caption{Countermeasure proposals based on TEE extensions.}
\label{t:tee-extensions}
\end{table}

The authors in~\cite{abdollahi2025earlyexperienceconfidentialcomputing} introduce a framework to assess the execution of ML models in ARM-based mobile devices. The main pillar of this work concerns the  evaluation of the Confidential Compute Architecture (CCA) technology with respect to the performance-privacy trade-offs. Using this approach, the authors extend the typical inference capabilities incorporated in existing TEE environments, towards a more efficient and secure region. In order to achieve that, CCA is designed using the Realm Management Extension (RME), a new component which introduces additional secure regions namely realms. This region is a TEE-like, special virtual machine, that is de-privileged, as it has only virtualised access to the system resources. Overall, the aim of the analysis is to evaluate whether CCA is sufficient countermeasure against data privacy attacks, including Membership Inference Attacks (MIA) and intellectual property violations.  

The authors of~\cite{moonasgard} propose ASGARD, a virtualisation-based TEE solution that can protect on-device DNNs on legacy Armv8-A SoCs. A significant security issue addressed in the paper is model stealing at the TEE layer. To counter this issue, ASGARD implements secure accelerator I/O passthrough, Trusted Computing Base reduction techniques, and exit-coalescing DNN execution planning, ensuring the security and integrity of the model inference process. The proposed solution is designed for Android devices.

The work in~\cite{10637594} highlights the limitations of existing TEE environments when multi-model execution is necessary. More specifically, the authors highlight the lack of multi-enclave solutions when more than one AI model are executed on-device. This stems from the fact that multiple applications on the same device that use AI models would need to utilise more than one enclave to protect themselves from model stealing and tampering attacks. Taking this as a basis, the authors explore the features of RISC-V implementations regarding multi-enclaves such as keystone and MultiZone.

The authors of~\cite{bertschi2024devloreextendingarmcca} propose Devlore, a system that protects various AI models, including LLMs and DL models, on ARM-based devices. A significant security issue addressed is model stealing and interrupt-based attacks, where an attacker may steal the model weights and user data at TEE layer. To counter this issue, Devlore implements a two-GPT design and interrupt isolation mechanisms, which ensure the security and integrity of the model inference process. The proposed solution is designed for devices that support ARM architecture, such as smartphones and embedded devices. 

The authors of~\cite{10.1145/3642970.3655845} propose GuaranTEE, a system that protects machine learning models, including DL models, on edge devices. A high risk addressed in the paper is model stealing and tampering, compromising the model weights and user data at the TEE layer. To counter this issue, GuaranTEE implements a framework that uses ARM's Confidential Computing Architecture (CCA) to provide attestable and private machine learning, which ensures the security and integrity of the model inference process. The proposed solution is designed for devices that support ARM architecture, such as smartphones and IoT devices. 

The work in~\cite{10.1145/3548606.3560627} enhances the GPU security of ARM-based endpoints to prevent sensitive info leakage inside the GPU. The work introduces StrongBox, the first TEE running on a GPU that is flexible, without targeting only secure inference, but rather it generalises to secure computation. This is achieved using isolated execution environments accessible only by the GPU. StrongBox leverages a smart way to reduce the runtime overhead by tagging specific GPU buffers as secure.

The authors in~\cite{RAHMAN2020103737} present a framework to secure QoS data on edge networks. More specifically, the authors propose a solution that safeguards the service composition process. The existing works in the literature rely on plaintext QoS data to assess which services need to be composed or which tasks and how they should be distributed to end-devices, causing privacy leaks. The proposed framework offers a QoS composition service that can be executed on encrypted data using Fully Homomorphic Encryption (FHE).




\section{Discussion}
\label{sec:discussion}

\begin{figure}

  \includegraphics[width=1\textwidth]{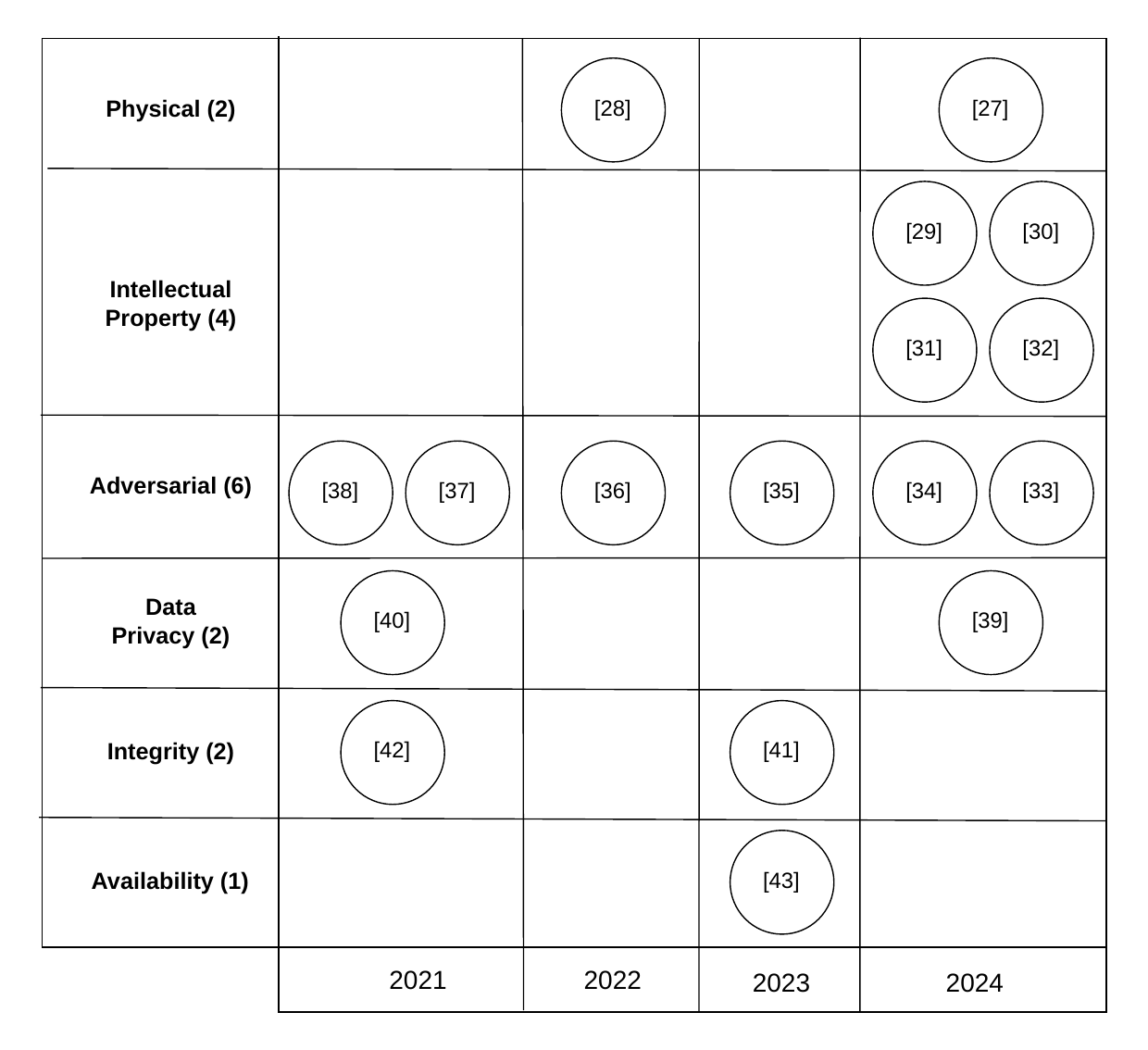}

  \caption{Attack taxonomy. 
  }
  \label{fig:AttackAnalysis}
\end{figure}

A survey of the attack literature shows common themes that currently define the threat environment. The works surveyed are presented chronologically in Figure~\ref{fig:AttackAnalysis}, sorted into the six threat categories defined in Section~\ref{sec:attack}. The attack category most explored is Adversarial with 6 works, of which 4 are Blackbox attacks, followed by Intellectual Property with 4. Physical, Data Privacy and Integrity attacks each contribute 2 works, while availability is the least explored category with only 1 study. From a temporal perspective, adversarial attacks are a category of continuous interest, since research on this topic has been consistently published from 2021 to 2024. On a year-by-year basis, 2024 has the highest concentration of publications, with a total of 8 works, reflecting the growing research interest in on-device AI security in recent years. Considering the entire period 2021-2024, it is clear that research concentrated on adversarial and intellectual property attacks, whereas other threat vectors affecting on-device AI have received relatively limited attention, with only a couple of studies addressing each of these categories. 

Table~\ref{t:attack-summary} summarises the effectiveness and scope of the attacks discussed in Section~\ref{sec:attack}. Physical attacks, including hardware and system-based attacks, are the most potent and reliable. GPU timing attacks, for example, achieved a perfect 100\% success rate. Model theft and intellectual property attacks achieved success rates between 33.83\% and 99\%. This often varies depending on the black-box query budgets and the complexity of target architectures. Adversarial attacks also vary widely in effectiveness (10.23\%-99.12\%), as their success depends mainly on query efficiency and perturbation budget. Privacy attacks report a narrow success range (92\%-93\%), which likely reflects limited coverage in the surveyed literature, not consistently high success rates. The above rate demonstrates the practical feasibility of tampering with deployed models. On the other hand, Integrity attacks such as DeepPayload, reached a success rate of 93.5\% but with a wide range, starting from 20\%. Finally, Availability attacks against end devices manage to increase battery consumption up to 21\% without sacrifising the model accuracy. 

\begin{table}
\centering

\small
\begin{tabular}{|>{\centering\arraybackslash}m{3cm}|>{\centering\arraybackslash}m{2cm}|>{\centering\arraybackslash}m{2.6cm}|>{\centering\arraybackslash}m{4cm}|}
\hline
\textbf{Category} & \textbf{Range} & \textbf{Peak} & \textbf{Observation} \\
\hline
Physical & 97.2-100\% & 100\% (GPU-timing) & Side-channels very effective \\
\hline
Intellectual Property & 33.83-99\% & 99\% (REDLC) & ML-based reconstruction dominates \\
\hline
Adversarial Attacks & 10.23-99.12\% & 99.12\% (GenDroid) & Query efficiency critical \\
\hline
Data Privacy & 92-93\% & 93\% (recall) & High effectiveness against FL \\
\hline
Integrity & 20-93.5\% & 93.5\% (DeepPayload) & Tampering highly successful \\
\hline
Availability & 14-21\%& 21\% (Sponge Poisoning) & Newer, more advanced mobile processors are more vulnerable to Sponge Poisoning\\
\hline
\end{tabular}
\caption{Attack effectiveness by category.}
\label{t:attack-summary}
\end{table}

\begin{figure}
  \includegraphics[width=1\textwidth]{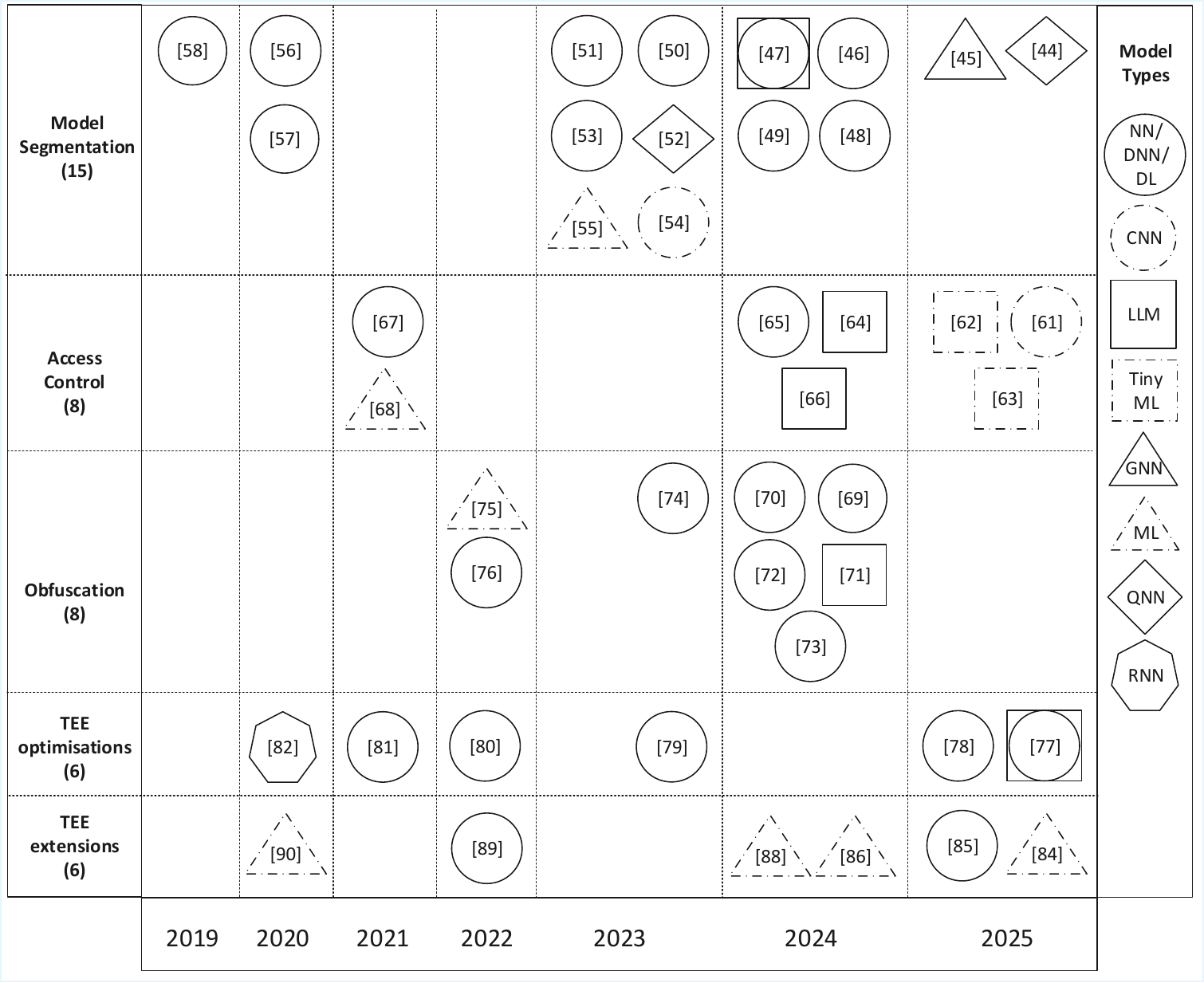}

  \caption{Protection mechanisms' taxonomy. Each shape represents the AI model that was examined in each work (circles: DNN/DL, dashed circles: CNN, rectangles: LLM, dashed rectangles: TinyML,  triangles: GNN, dashed triangles: ML, diamonds: QNN, heptagons: RNN).}
  \label{fig:DefenseAnalysis}
\end{figure}

Against this threat landscape, Figure~\ref{fig:DefenseAnalysis} provides an overview of the core defence strategy employed in each study, organising defence solutions in chronological order. Every work is illustrated using a different shape based on the fundamental AI category it explores. In this respect, circles depict works that discuss countermeasures on DNN/DL deployments and dashed circles CNN models. Rectangles highlight research that investigate protection measures against attacks on LLMs, while dashed rectangles indicate solutions for TinyML. Triangles depict GNN-related articles and dashed triangles works that protect ML models. In contrast, diamonds identify publications that investigate QNN-based systems, and finally, heptagons denote RNNs.
On a year-by-year basis, the research interest into on-device inference protection increased from 1 paper in 2019, to 3 to 4 papers per year between 2020 and 2022, and more than 8 papers per year between 2023 and 2025, with a peak of 14 works in 2024.
Regarding the classification of defences, model segmentation emerges as a prominent approach, with 15 works spanning the period from 2019 to 2025. The majority of the studies in this category was published from 2023 to 2025, whereas only three works were published in 2019-2020. In contrast, the access control domain is maturing at a slower pace, with only 2 works in 2021, and with 3 studies in each of the years 2024 and 2025. The obfuscation category has experienced an equal growth in terms of numbers with the access control category, but with an uneven distribution across the considered years. More specifically, 2 works were published in 2022, 1 in 2023, and the remaining 5 works were published in 2024. Finally, the TEE domains have contributed a moderate number of studies. Specifically, TEE optimisations yielded a total of 6 works, with almost one per year from 2020 to 2025. Similarly, the TEE extensions category comprises 2 works from 2020 to 2022, 2 works in 2024, and 2 works in 2025.

An additional pillar that shapes the map of defence solutions is the type of AI model being protected. As Figure~\ref{fig:DefenseAnalysis} illustrates, the majority of papers focus on defending DNN/DL deployments, denoted by circles. In total, 25 solutions were published between 2019 and 2025 in this category, with a notable peak of 9 publications in 2024. In particular, DNN/DL is the only model type represented across all attack categories and examined years. In contrast, solutions addressing LLMs, marked as rectangles, are relatively scarce, with only 5 publications exclusively in 2024 and 2025, reflecting the recent emergence of this research area. For TinyML, only 2 articles were published in 2025 that focused solely on integrity. Work on QNNs and CNNs has been limited to 2 articles each, both of which address intellectual property concerns in 2023 and 2025. RNNs and GNNs have each been covered by a single publication, appearing in 2020 and 2025, respectively, with distinct research directions. 
Both address data privacy considerations.

Overall, the analysis of defence solutions proves that the research interest is mostly gathered around DL/DNN approaches. This shift towards DL/DNN can be justified from the wide adoption of these deployments in real-world systems. Regarding the other AI paradigms, the reduced number of studies signifies that the specific areas are quite immature, leaving open space for new contributions. Nevertheless, the research community in this area seems also to consider other model types than DL/DNN, as seen by the emerge of works protecting LLMs (5 works in 2024-2025) and TinyML models (2 works in 2025). As the weaker AI categories attract more attention, we expect a higher number of defence contributions in these categories over the next few years.

\begin{figure}

  \includegraphics[width=1\textwidth]{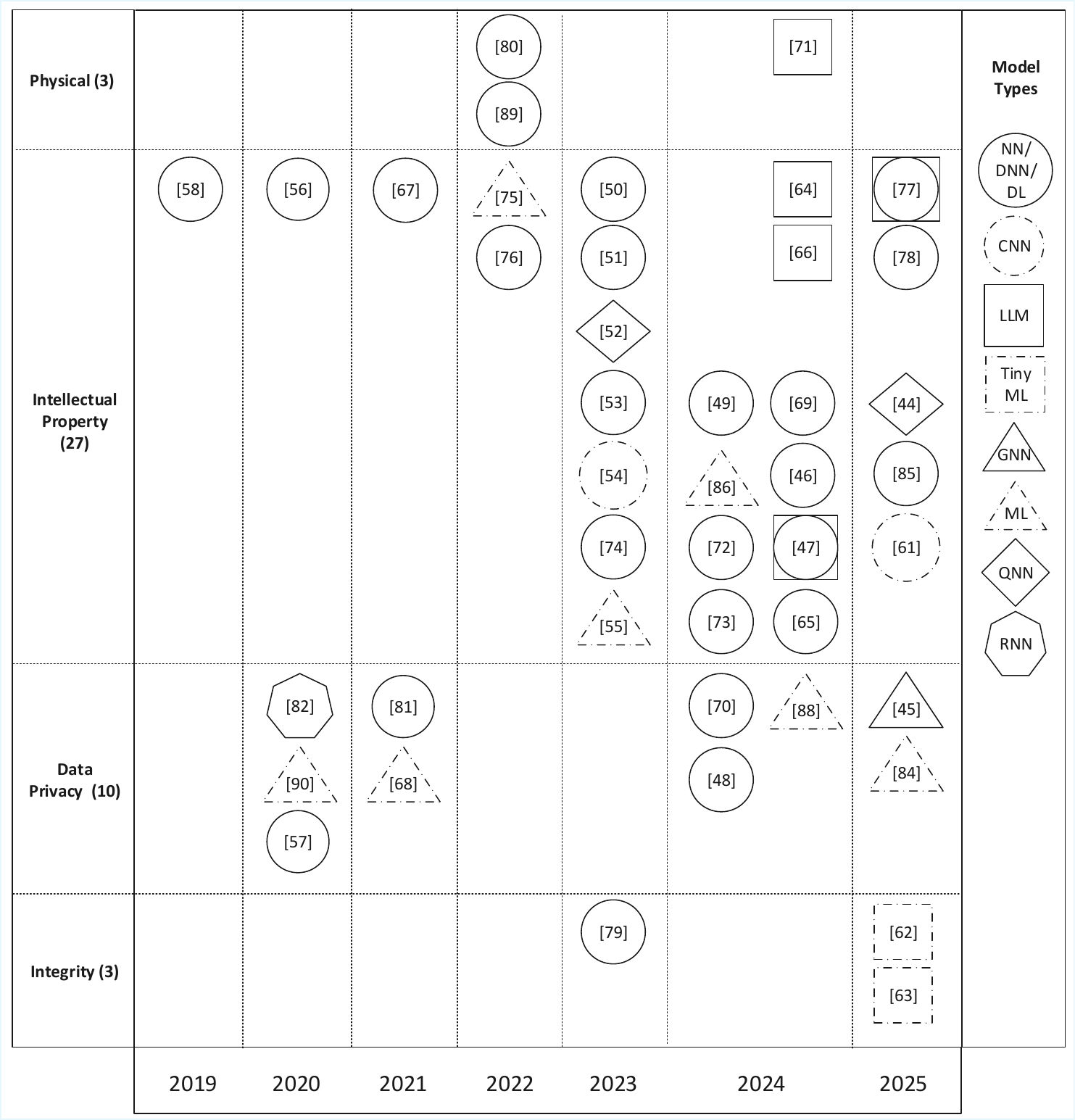}

  \caption{Mapping of defences to attack categories.}
  \label{fig:DefenseAnalysisCategories}
\end{figure}

A further dimension to consider in this analysis is the category of attacks addressed by each protection mechanism; Figure~\ref{fig:DefenseAnalysisCategories} depicts a mapping of protection mechanisms to attack categories.
Notably, the intellectual property and data privacy categories attract significant attention within the research community, as evidenced by the substantial number of studies dedicated to these areas, with 27 and 10 research works, respectively. By contrast, adversarial and availability threats do not appear in the figure, as we found no defence papers that explicitly mitigate these attacks. Only three studies address integrity threats, and a further three protect against physical attacks. 
Finally, the distribution by publication year shows that 14 defence papers were published in 2024, followed by 8 in 2023 and 9 in 2025.

\section{Challenges \& future directions}
\label{challenges}

With AI models blooming on a daily basis, edge and mobile devices are becoming the new paradigm to deliver privacy-preserving AI services. This paradigm offers significant advantages for certain, regulated industries that need to meet strict security and privacy requirements stipulated by regulation, such as the Cyber Resilience Act (CRA)~\cite{cra2024}, the General Data Protection Regulation (GDPR)~\cite{gdpr}, and the EU Artificial Intelligence Act (AI Act)~\cite{euaiact2024}, 
which often prohibit the transfer of sensitive customer data to the cloud and aim to ensure the safe and responsible development of AI. By leveraging on-device AI inference, companies can effectively address privacy-sensitive market segments, opening new business opportunities. However, this approach introduces also a set of challenges, mainly related to the inherent constraints of mobile devices, such as limited computational power and energy storage, as well as security vulnerabilities.

One of the primary challenges in on-device AI inference is the optimisation of models for resource-constrained devices. Model compression techniques, including pruning and quantisation, are essential for deploying large models on edge devices. However, these techniques often involve trade-offs between model size and accuracy, requiring careful consideration of the specific application requirements. For example, the size of the tiny version of the Whisper model for client-side transcription is around 40MB~\cite{whisper-tiny-openai}. This optimisation has implications on the inference result as the accuracy remains low, highlighting the need for more advanced model optimisation techniques. Linking this limitation to on-device inference security, model compression can potentially introduce security vulnerabilities in the model, by creating new attack surfaces or making the model more susceptible to attacks, including adversarial~\cite{lin2019defensive} or model inversion attacks~\cite{brown2024does}.

Resource limitations could also have an impact on the availability of security mechanisms on resource-constrained devices. 
This implies that deploying security mechanisms such as encryption or intrusion detection systems may be constrained or weakened. In particular, state-of-the-art techniques (e.g., homomorphic encryption or differential privacy) may be impractical to run effectively, leaving the device more vulnerable to attacks.

Another challenge is the diversity of edge devices, which makes it difficult to develop a common solution that can operate on devices with different computational capabilities, memory constraints, and power consumption characteristics. For example, a different data protection solution is needed in a laptop compared to an RFID tag. Shielding edge devices with robust security mechanisms is not always possible, but other approaches can be adopted. For example, some solutions on this front concern hardware acceleration or other hybrid methods, such as differential privacy and perturbation~\cite{10601684}. Heterogeneous computing architectures that combine CPUs, GPUs, and Neural Processing Units (NPUs) are critical for balancing power-performance efficiency in on-device AI. However, the development of efficient accelerators that can adapt to different performance, power, and cost constraints remains an open challenge.

Overall, one direction to protect AI deployments on edge devices concerns the use of the defence mechanisms that have been analysed in Section~\ref{sec:defence}. 
Another major direction is improving model design for efficient edge inference. Future research should focus on developing architectures that are specifically designed for edge deployment, balancing accuracy, efficiency, robustness and security. To this end, Neural Architecture Search (NAS) can automate the discovery of compact model architectures optimised for specific edge devices~\cite{Cai_2022}. Moreover, hardware-aware NAS considers the target hardware characteristics during the search process, enabling the discovery of architectures that achieve optimal performance on specific devices. Efficient models can enhance privacy by making on-device inference more feasible. At the same time, future research could consider hardware-aware NAS with multi-objective optimisation that factors in robustness against adversarial attacks alongside hardware efficiency, creating more secure models for real-world deployment~\cite{benmeziane2021comprehensive}.


One promising technique that could contribute to privacy protection is unlearning. It enables on-device AI models to forget or remove the influence of specific training data or experiences, which can be particularly useful in scenarios where data is no longer relevant, has been compromised, or needs to be deleted for privacy or regulatory reasons. With unlearning, models can be updated to remove the impact of sensitive or outdated data, improving data privacy and reducing the risk of model inversion attacks. Moreover, unlearning can help organisations comply with regulations such as GDPR, which requires that personal data be deleted upon request. However, developing effective unlearning techniques that can balance data removal with model performance is a challenging task, and ensuring that data is truly forgotten and not retained in any form is crucial to prevent potential security risks. As a future direction, researchers and developers can explore various unlearning algorithms and techniques, such as differential privacy and secure data deletion, to improve the security and privacy of on-device AI inference and enable the development of more trustworthy and compliant AI models.

On the regulatory front, future research should contribute to the development of standards for on-device AI, including standards for performance measurement, security evaluation, and interoperability. Standardised approaches will facilitate the comparison of different solutions and enable more efficient system integration. Moreover, 
regulatory compliance tools and techniques should be developed to help AI developers meet the requirements of emerging regulations. This includes tools for privacy impact assessment and explainability that can demonstrate compliance without compromising privacy.

\section{Conclusions}
\label{sec:conclusions}

On-device inference has rapidly moved from a highly-specialised domain to a mainstream design choice for privacy-sensitive and latency-critical applications. Given that existing review works lack a detailed analysis of the security aspects of on-device AI inference, this paper presents a survey of the associated security threats and defensive mechanisms on edge and mobile devices. To better organise and present the surveyed literature, we propose taxonomies of both attacks and defences, and additionally organise the analysed works chronologically by publication year.

Our results have shown that the resulting threat landscape is both broad and practically exploitable: across 2019–2025, published attacks demonstrate high success rates, with at least one attack having 93\% effectiveness or more, across all attack categories. Moreover, nearly half of the surveyed attack papers were published in 2024, with the remainder appearing between 2021 and 2023, indicating a recent rise in research interest on the topic. The most actively researched attack vector is adversarial threats, covering 35\% of the 17 reviewed publications. 

Against this background, the defence literature remains unevenly distributed.
Regarding the strategies used by inference protection mechanisms, around one third of the studied solutions rely on model segmentation, whereas there is a balance in the usage of other defence techniques. 
A key observation emerges when mapping defences to the threats they address. Our mapping of 44 defence papers indicates that around two thirds prioritise intellectual property protection, even though intellectual property attacks account for only about one quarter of the surveyed attack papers. By contrast, we found no defence papers addressing availability and, notably, none targeting adversarial attacks, despite the later comprising roughly one third of the attack literature. This highlights critical gaps between existing attacks and available mitigations, leaving space for further research in the development of on-device inference protection mechanisms against known attacks.

Overall, our findings indicate that securing on-device inference requires a shift from protections targeting mainly intellectual property to a more holistic approach that considers the special characteristics of on-device AI model deployment, such as physical exposure, heterogeneous setups, and hardware limitations. Moreover, appropriate defence designs are needed that jointly optimise security, accuracy, and performance. Future research towards trustworthy on-device AI systems can build on the present survey, which identifies significant mismatches between well-known attack vectors and underdeveloped or missing mitigations.





\section*{CRediT authorship contribution statement}
ZT: Conceptualization, Investigation, Resources, Writing - Original Draft, Visualization, Project administration; AF: Conceptualization, Methodology, Investigation, Writing - Original Draft, Visualization; GK: Validation, Investigation, Writing - Review \& Editing, Supervision, Project administration; VK: Investigation, Writing - Original Draft; MA: Investigation, Supervision.

\bibliographystyle{elsarticle-num}
\bibliography{cas-refs}
\end{document}